\documentstyle[12pt]{article}
\def\beq{\begin{equation}}
\def\eeq{\end{equation}}
\def\bea{\begin{eqnarray}}
\def\eea{\end{eqnarray}}
\def\nn{\nonumber}
\def\ba{\begin{array}}
\def\ea{\end{array}}
\def\d{\partial}
\def\v{\vert}

\def\r{\rangle}
\def\one{1\hskip -1mm{\rm l}}

\begin{document}
% \begin{flushright} hep-th/9701097
% \end{flushright}
%\rightline{TIFR/TH/97-01}
%\rightline{January 1997}
\baselineskip16pt
\smallskip
\begin{center}
{\large \bf \sf
Jost solutions and quantum conserved quantities \\
of an integrable derivative nonlinear 
      Schr${\rm {\ddot o}}$dinger model }

\vspace{1.3cm}

{\sf B. Basu-Mallick\footnote{ 
e-mail address: biru@theory.saha.ernet.in }
and Tanaya Bhattacharyya\footnote{e-mail address:
tanaya@theory.saha.ernet.in } },

\bigskip

{\em Theory Group, \\
Saha Institute of Nuclear Physics, \\
1/AF Bidhan Nagar, Kolkata 700 064, India } \\
\bigskip

\end{center}

\vspace {1.1 cm}
\baselineskip=17pt
\noindent {\bf Abstract }

We study differential and integral relations for the quantum 
Jost solutions associated with an integrable derivative nonlinear 
Schr${\rm {\ddot o}}$dinger (DNLS) model. 
By using commutation relations between such Jost solutions and
the basic field operators of DNLS model, we explicitly construct 
first few quantum conserved quantities of this system including 
its Hamiltonian. It turns out that this quantum Hamiltonian has a 
new kind of coupling constant which is quite different from the 
classical one.  This modified coupling constant 
plays a crucial role in our comparison between
the results of algebraic and coordinate 
Bethe ansatz for the case of DNLS model. We also find out the range 
of modified coupling constant for which the quantum $N$-soliton state 
of DNLS model has a positive binding energy.

\bigskip

\baselineskip=16pt
\vspace {.6 cm}
\noindent PACS No. : 11.10.Lm , 11.30.-j, 02.30.Ik, 03.65.Fd 

\vspace {.1 cm}
\noindent Keywords : Derivative nonlinear 
Schr${\rm {\ddot o}}$dinger model, Jost solution, Yang-Baxter equation, 
 Algebraic Bethe ansatz

\newpage

\baselineskip=17pt
\noindent \section {Introduction }
\renewcommand{\theequation}{1.{\arabic{equation}}}
\setcounter{equation}{0}

\medskip

Conserved quantities associated with quantum integrable models
in low dimensions have recently found interesting applications 
in many topics of physics like exact calculations of transport 
properties in mesoscopic electronic devices and distribution of 
energy level spacing in quantum chaotic systems [1,2]. In the 
framework of quantum inverse scattering method (QISM), one can 
formally generate such conserved quantities by expanding 
the trace of monodromy matrix 
in a power series of spectral parameter [3-6]. The
Lax operator associated with monodromy
matrix satisfies quantum Yang-Baxter equation (QYBE). As a result,
all of these quantum conserved quantities commute among themselves. 
Thus, for constructing a quantum integrable 
field model or spin chain, it is natural to start with a suitable
quantum Lax operator which satisfies QYBE 
and find out corresponding 
conserved quantities including the Hamiltonian.

However, explicit construction of these conserved quantities in terms 
of basic quantum field or spin operators often turns out to be a challenging
 task which has inspired the application of several ingenious techniques. 
For example, in the case of one dimensional quantum integrable spin chains 
like Heisenberg model, supersymmetric t-J model and Hubbard model, 
one can explicitly construct the conserved quantities 
in a recursive way by using appropriate `ladder operators' [7-10]. 
While dealing with $1+1$-dimensional classically integrable 
 field theoretical systems like nonlinear 
Schr${\rm {\ddot o}}$dinger (NLS) model,
it is again possible to explicitly construct the conserved quantities 
in a recursive way by solving corresponding 
 Riccati equations.
However this recursive method of finding the conserved quantities 
does not usually work for quantum integrable field models, where
 the presence of normal ordering might lead to non-uniformness in the 
 asymptotic expansion of monodromy matrix in powers 
of spectral parameter. As a result, it may not be possible to obtain
 all quantum conserved quantities simply as normal ordered versions of the 
corresponding classical conserved quantities [11-13].  
Fortunately, however, this problem does not occur for the case of some lower 
conserved quantities of quantum NLS model, which are
generated by first few terms 
in the asymptotic expansion of monodromy matrix [14]. Consequently,
conserved quantities associated with number of particles, 
momentum as well as 
Hamiltonian of the quantum NLS model can be 
obtained just as normal ordered versions of the  
corresponding classical conserved quantities.
The Hamiltonian of quantum integrable Sine-Gordon 
model can also be obtained in a similar way 
from the corresponding classical Hamiltonian [3,5].

Even though the Hamiltonians of quantum integrable 
field models usually coincide with the normal ordered versions of 
the corresponding classical Hamiltonians, there is no guarantee that
this thumb rule will always be obeyed.
The main purpose of the present article is to construct the 
quantum Hamiltonian of a derivative 
nonlinear Schr${\rm {\ddot o}}$dinger (DNLS) model 
through the corresponding Lax operator
and explore how this quantum Hamiltonian is related to its classical 
counterpart.  In this context it may be noted that there exist 
two variants of classically integrable 
DNLS model in $1+1$-dimension [15,16], which have found 
applications in physical systems like circularly polarized nonlinear 
Alfven waves in a plasma [17,18]. However, only one among these variants of
  DNLS model is known to be associated with an ultralocal Poisson
Bracket (PB) structure which is very suitable for quantization
through QISM [19,20]. 
 The equation of motion for such classical DNLS model is given by [16]
\beq
i \d_t \psi (x,t) + \d_{xx}\psi(x,t) - 4i \, \xi \, 
{\psi^* (x,t)}{\psi (x,t)} 
\d_x{\psi(x,t)} \, = \,  0 \, ,
\label {a1}
\eeq 
where 
$\d_t \equiv {\d \over \d t}$,
$\d_x \equiv {\d \over \d x}$,
$\d_{xx} \equiv \frac{\d^2}{\d x^2}$ and  
$\xi$ is a real parameter representing the 
strength of the nonlinear interaction term.
The Lax operator related to this DNLS model may be written in the form
[19,21]
\bea
       U( x, \lambda ) =
i \pmatrix { \xi\psi^*( x )\psi( x ) - {\lambda^2}/4  &
\xi\lambda\psi^*( x ) \cr
\lambda\psi( x ) &  -\xi\psi^*( x )\psi( x ) + \lambda^2/4 } \, ,
\label{a2}
\eea
where $\lambda$ denotes the spectral parameter
and $\psi (x) , \, \psi^* (x) $ represent 
 field variables at some fixed time (which is suppressed here
and all along in the following).
By solving the Riccati equation associated with Lax operator (\ref {a2}), 
one can explicitly construct the conserved quantities
 for this DNLS model in a recursive way.
The first few among such 
infinite number of classical conserved quantities,
 representing the mass, momentum and Hamiltonian of the DNLS
system respectively, are given by [19]
\bea
~~~~~~~~~~~~&& N = \int_{-\infty}^{+\infty}\psi^*(x)\psi(x) \,  dx \, , ~~
P = -i \, \int_{-\infty}^{+\infty}\psi^* (x) \d_x \psi (x) \, dx \, , \nn  
~~~~~~~~~~~ (1.3a,b) \\
~~~~~~~~~~~~~~&& H = \int_{-\infty}^{+\infty} 
\left\{ \, -  \psi^*(x) \, \d_{xx} \psi(x) \, + \, i\xi \,
 {\psi^*}^2(x) \, \d_x  \psi^2(x)   \, \right \}
 dx \, . \nn ~~~~~~~~~~~~~~~~~~ (1.3c)
\eea
\addtocounter {equation}{1}
The field variables appearing in the Lax operator (\ref {a2})
 obey the following equal time 
  PB structure: $\{ \psi(x),
\psi(y) \} = \{ \psi^*(x), \psi^*(y) \} = 0$,  $\{ \psi(x), \psi^*(y) \} =
-i \delta(x-y)$.  
With the help of this ultralocal PB structure, it can be shown that 
the Lax operator (\ref {a2}) satisfies classical Yang-Baxter
equation. As a result, infinite number of conserved quantities associated 
with DNLS model (\ref {a1}) yield vanishing PB relations 
among themselves [19]. 
This fact establishes the classical integrability 
of DNLS model (\ref{a1}) in the Liouville sense. 

It is remarkable that the 
 integrability property of the above mentioned classical DNLS model 
 can be preserved even after quantization. In this quantized version of
DNLS model, the basic field operators 
satisfy equal time commutation relations given by
\beq
\Big[\psi( x ), \psi( y )\Big] = 
\left[ \psi^\dagger( x ) , \psi^\dagger( y ) \right] = 0 ,~~~~
\left[ \psi( x ) , \psi^\dagger( y ) \right] = \hbar\delta( x - y ) \, ,
\label {a4}
\eeq
$\hbar $ being the Planck's constant. The corresponding  
 vacuum state is defined through the relation: $\psi(x) \v 0 \r = 0$.
The most natural way of constructing such quantum 
integrable DNLS model, possessing infinite number of mutually 
commuting conserved quantities, is to first find out the 
  quantum analogue of classical Lax operator (\ref {a2})
 which would satisfy the QYBE.
However, it can be easily shown that QYBE is not satisfied 
if the normal ordered version of 
classical Lax operator (\ref {a2}) is directly chosen 
as the quantum Lax operator of DNLS model.
The correct form of this quantum Lax operator, satisfying QYBE in continuum,
is given by [21]
\bea
       {\cal U}_q ( x, \lambda ) =
i \pmatrix { f \, \psi^\dagger(x)\psi( x )
 - {\lambda^2}/4  &
\xi\lambda\psi^{\dagger}( x ) \cr
\lambda\psi( x ) &  - g \, \psi^\dagger(x)\psi(x)
 + \lambda^2/4 } \, ,
\label {a5}
\eea 
where $f = \frac{\xi
e^{-i\alpha/2}}{\cos \alpha / 2}$ , $g = \frac{\xi
e^{i\alpha/2}}{\cos \alpha / 2}$ and $\alpha$ 
a real parameter ($-\frac{\pi}{2} < \alpha \leq \frac{\pi}{2}$) 
which is uniquely determined through the relation 
\bea
\sin\alpha = -\hbar\xi \, .
\label {a6}
\eea
Thus the quantum Lax operator (\ref {a5}) depends not only
 on the parameter $\xi$, but also on the Planck's 
constant $\hbar $. It is clear from eqn.(\ref {a6}) that, for any fixed value 
of $\xi $, $\alpha \rightarrow 0$ limit is essentially equivalent to
$\hbar \rightarrow 0$  limit. Since 
 $f\rightarrow \xi$ and $g\rightarrow \xi$ at $\alpha \rightarrow 0$ limit,
the quantum Lax operator (\ref {a5}) reproduces the classical
Lax operator (\ref {a2}) at $\hbar \rightarrow 0$  limit. 
With the help of Lax operator (\ref {a5}) or its lattice version 
[19,20], one can easily construct the monodromy matrix of quantum DNLS model
in continuum. Quantum conserved quantities can be defined formally
through the diagonal elements of this monodromy matrix, by expanding them 
in the power series of spectral parameter.
By applying algebraic Bethe ansatz to such formally defined
quantum conserved quantities, 
one can  derive their  exact eigenfunctions and eigenvalues
 for scattering as well as 
bound soliton states [19,21]. Moreover, one can also construct the reflection 
operators for the DNLS model satisfying the Zamolodchikov-Faddeev algebra
and find out the $S$-matrix for the two body scattering [21]. 

In spite of these studies on quantum DNLS model,
the problem of explicitly constructing its conserved 
quantities in terms of 
 basic field operators like $\psi (x)$ and $\psi^\dagger(x)$ 
has not been addressed so far. In particular it is not known 
whether, in analogy with the quantum NLS model and sine-Gordon model, 
the Hamiltonian of quantum DNLS model can also be obtained as the normal 
ordered version of the corresponding classical Hamiltonian (1.3c). 
The explicit form of such quantum Hamiltonian
 would clearly play a central role in interpreting various properties of this 
 field model in the language of associated 
quantum mechanical many-particle system. 
In this context it should be observed that, if the normal
ordered version of classical Hamiltonian (1.3c) is projected on an 
$N$-particle Hilbert space, that would yield an $N$-particle bosonic system 
  interacting through the derivative $\delta$-function
potential [22,23], where $\xi$ represents
 the strength of the interaction.
 Equation (\ref {a6}) however imposes a restriction on the value of 
this coupling constant as $\v \xi \v \leq {1\over \hbar}$. 
Thus it is evident that, if the 
normal ordered version of the classical Hamiltonian (1.3c) represents the 
quantum Hamiltonian of DNLS model, the corresponding
 $N$-particle bosonic system can not be solved 
through QISM for $ \v \xi \v  > {1\over \hbar}$. 
On the other hand, it is known that this $N$-particle bosonic system 
  with derivative $\delta$-function interaction can be solved exactly
for any value of its coupling constant 
through the coordinate Bethe ansatz [22-24]. 
Thus one faces a rather curious limitation about the 
 applicability of algebraic Bethe ansatz to the case of quantum
DNLS model.

It is clear that,  some direct  method of finding the explicit form of 
 quantum Hamiltonian associated with 
the Lax operator (\ref {a5}) of DNLS model may help us to  
 resolve the above mentioned problem. In this context,
we recall a work by Case [11] where
 first few conserved quantities of the quantum NLS model are explicitly 
constructed and their spectra are also derived in the following way. 
At first, Jost solutions associated with the Lax operator of 
quantum NLS system are considered.  The scattering data, i.e. 
 elements of monodromy matrix, are identified with the 
 Wronskians corresponding to these Jost solutions. Subsequently 
it is proposed that the commutators between 
quantum conserved quantities of the NLS model and Wronskians 
obey the so called `fundamental relation'. 
This relation can generate the spectra 
of all quantum conserved quantities in an algebraic way.  
The explicit form of the 
 first few quantum conserved quantities of NLS model
 are obtained from the requirement of 
satisfying this fundamental relation.  

The above mentioned way of constructing quantum conserved quantities and 
finding their spectra is clearly different from the usual
algebraic Bethe ansatz in QISM.  However, 
in complete analogy with QISM, finding an appropriate quantum Lax operator 
is the starting point of Case's approach. So this approach
gives us valuable insight about the explicit form of quantum 
conserved quantities which can be obtained from 
 the trace of monodromy matrix in QISM. 
In this article we shall study quantum DNLS model through this 
approach which is complimentary to QISM. In Section 2, 
 we briefly recapitulate the construction of quantum Lax operator of 
DNLS model through a variant of QISM 
which is directly applicable to field theoretical systems and 
 also discuss how the related conserved quantities 
can be diagonalised through algebraic Bethe ansatz [21]. 
In Section 3 we use the quantum Lax operator and monodromy matrix,
obtained through QISM, for defining the 
Jost solutions of DNLS model. It is surprisingly found that, 
in contrast to the case of NLS model, differential equations satisfied by
 Jost solutions associated with
boundary conditions at $x \rightarrow \infty $ 
and $x \rightarrow - \infty $ do not coincide with each other.
%Wronskians defined in terms of such Jost solutions represent 
%the elements of monodromy matrix for the DNLS model. 
Using the Wronskians
and some other bilinear functions of these Jost solutions, 
in Section 4 we propose the `fundamental relation' for the DNLS model 
and derive the spectra for all conserved 
quantities which would satisfy this relation.
Here we also discuss how the conserved quantities satisfying the
above relation are related to the conserved quantities 
which are formally defined in the framework of QISM. 
In Section 5, we discuss about the 
necessary tools for finding out the explicit form of conserved quantities 
satisfying the fundamental relation. In particular, 
we derive the commutation relations between the 
Wronskians and basic field operators of the system. 
In Section 6, we construct the explicit form of first few conserved quantities  
of the quantum DNLS model including its Hamiltonian. Interestingly, it is 
 found that the interaction part of this quantum Hamiltonian has a 
new kind of coupling constant which is quite different from the classical one. 
Here we also derive the condition on this coupling constant
for which the quantum $N$-soliton state of DNLS model has a positive
binding energy. Section 7 is the concluding section.

\vspace{1cm}

\noindent \section {Application of QISM to DNLS model}
\renewcommand{\theequation}{2.{\arabic{equation}}}
\setcounter{equation}{0}

\medskip

As mentioned earlier,
the monodromy matrix plays a key role in formally
 generating the quantum conserved quantities of DNLS model 
and in diagonalising those conserved quantities through QISM. 
With the help of Lax operator (\ref {a5}), 
one can define the quantum monodromy matrix of DNLS model 
on a finite interval as 
\beq
{\cal T} ^{x_2}_{x_1}(\lambda) = \quad  :{\cal P} \exp \int_{x_1}^{x_2} {\cal
U}_q(x,\lambda) dx : \, ,
\label{b1}
\eeq
where ${\cal P}$ denotes the path ordering and the symbol 
  $::$ denotes the normal ordering of operators. It is 
evident that this monodromy matrix satisfies differential equations of the
form
\bea
~~~~~~\frac{\partial}{\partial x_2}{\cal T}^{x_2}_{x_1}( \lambda )  \, = \, 
 : {\cal U}_q( x_2 ,
\lambda ){\cal T}^{x_2}_{x_1}( \lambda ) : \,  , ~~~~
\frac{\partial}{\partial x_1}{\cal T}^{x_2}_{x_1}( \lambda ) \, = \,  -
 : {\cal T}^{x_2}_{x_1}(\lambda) {\cal U}_q(x_1 ,\lambda) : \, . \nn
 \, ~~~(2.2a,b)
\eea
\addtocounter{equation}{1}
By using these differential equations
and canonical commutation relations (\ref {a4}),
 it can be shown that the 
direct product of two such quantum monodromy matrices satisfies QYBE
given by [21]
\beq
R( \lambda, \mu ){\cal T}^{x_2}_{x_1}( \lambda ) \otimes 
{\cal T}^{x_2}_{x_1}( \mu ) = {\cal T}^{x_2}_{x_1}( \mu ) \otimes 
{\cal T}^{x_2}_{x_1}( \lambda )R( \lambda, \mu ) \, .
\label{b3}
\eeq
Here $R(\lambda, \mu)$  is a
 $(4\times 4)$ matrix with $c$-number elements like 
\beq
R( \lambda, \mu ) = \pmatrix { 1 & 0 & 0 & 0 \cr
0 & s( \lambda, \mu ) & t( \lambda, \mu ) & 0 \cr
0 & t( \lambda, \mu ) & s( \lambda, \mu ) & 0 \cr
0 & 0 & 0 & 1} \, ,
\label{b4}
\eeq
with $t(\lambda, \mu)=\frac{\lambda^2 - \mu^2}{\lambda^2 q - \mu^2
q^{-1}} , ~ s( \lambda, \mu ) = \frac{( q - q^{-1} )\lambda\mu}
{\lambda^2 q - \mu^2 q^{-1}} $ and $ q = e^{-i\alpha} $. It is 
mentioned earlier that the real parameter $\alpha $, which is present 
both in Lax operator (\ref {a5}) and $R$-matrix (\ref {b4}),
is fixed through the relation (\ref {a6}).
Consequently, QISM is applicable for quantum 
DNLS model when the parameter $\xi $ satisfies a restriction 
given by $\v \xi \v \leq {1\over \hbar}$. 

Next, by using the expression of ${\cal T}^{x_2}_{x_1}( \lambda )$ in
(\ref{b1}), we define the quantum monodromy matrix 
on an infinite interval limit as
\beq
{\cal T}(\lambda) = 
\lim_{\stackrel {x_2 \rightarrow + \infty} {x_1 \rightarrow -\infty}} 
e(-x_2,\lambda) {\cal T}^{x_2}_{x_1}(\lambda)e(x_1,\lambda) =
{\cal T}_+( x, \lambda ) {\cal T}_-( x, \lambda )  \, ,
\label {b5}
\eeq
where $e( x, \lambda ) = e^{-\frac{i\lambda^2x}{4}\sigma_3 }$ and 
\bea
~~~~~~~ {\cal T}_+( x, \lambda ) = \lim_{x_2 \rightarrow + \infty} 
e(-x_2,\lambda) {\cal T}^{x_2}_x(\lambda)  \, ,~~
{\cal T}_-( x, \lambda ) = \lim_{x_1 \rightarrow -\infty} 
{\cal T}^x_{x_1}(\lambda)e(x_1,\lambda) \, .  ~~~~~~~~
(2.6a,b) \nn 
\eea
\addtocounter{equation}{1}
Taking into account that the quantum Lax operator (\ref{a5}) obeys 
certain symmetry properties [21] and assuming
$\lambda$ to be a real parameter, 
one can express the quantum monodromy
matrix (\ref {b5}) in a symmetric form given by
\bea
{\cal T}(\lambda)=\pmatrix {A(\lambda) & -\xi B^\dagger(\lambda) \cr
                          B(\lambda) & A^\dagger(\lambda)} \, ,
\label {b7}
\eea
and find that these operator valued elements satisfy
 relations like  $A(-\lambda)= A(\lambda), ~
B(-\lambda)= - B(\lambda)$. Moreover, 
 it is easy to show that these elements act on 
the vacuum state as: 
$A(\lambda) \v 0 \r = \v 0 \r, ~B(\lambda) \v 0 \r = 0$.
With the help of eqns.(\ref {b3}) and (\ref {b5}),
one may now obtain QYBE for the quantum monodromy matrix on an 
 infinite interval as [21]
\beq
R( \lambda, \mu ) C_+( \lambda, \mu ) {\cal T}( \lambda ) 
\otimes {\cal T}( \mu ) C_-( \lambda, \mu ) = 
C_+( \mu, \lambda ) {\cal T}( \mu ) \otimes {\cal T}(\lambda)
 C_-( \mu, \lambda ) R( \lambda, \mu ) \, ,
\label{b8}
\eeq
where
\bea
C_+( \lambda, \mu ) = \pmatrix { 1 & 0 & 0 & 0 \cr
0 & 1 & 0 & 0 \cr
0 & \rho_+( \lambda, \mu ) & 1 & 0 \cr
0 & 0 & 0 & 1} \, ,~~
C_-( \lambda, \mu ) = \pmatrix { 1 & 0 & 0 & 0 \cr
0 & 1 & 0 & 0 \cr
0 & \rho_-( \lambda, \mu ) & 1 & 0 \cr                                 
0 & 0 & 0 & 1}  \, ,
\label {b9}
\eea
and
$$
\rho_\pm( \lambda, \mu ) = \mp \frac{2i\hbar \xi  \lambda \mu}
{\lambda^2 - \mu^2}
+ 2 \pi \hbar \xi  \lambda \mu \delta( \lambda^2 - \mu^2 ) 
= \mp \frac{2i\hbar\xi 
\lambda \mu}{\lambda^2 - \mu^2 \mp i\epsilon} \, .
$$
By inserting the explicit expressions for $R( \lambda, \mu )$ (\ref{b4}),
$C_{\pm}(\lambda,\mu)$ (\ref {b9}) and 
${\cal T}( \lambda )$ (\ref{b7}) to QYBE (\ref{b8}) and comparing its
matrix elements from both sides, we finally obtain
\bea
&&~~~~\Big[ A( \lambda ), A( \mu ) \Big] = 0 \, , ~~
\left[ A( \lambda ), A^\dagger( \mu ) \right] = 0 \, ,~~
\Big[ B( \lambda ), B( \mu ) \Big] = 0 \, , \, ~~ \,  \nn 
~~~~~~~~~~~~(2.10a,b,c) \\
&&~~~~A(\lambda) B^\dagger(\mu) = \frac{\mu^2 q - \lambda^2 q^{-1}}{\mu^2 -
\lambda^2 - i \epsilon} B^\dagger( \mu ) A( \lambda ) \, , ~\nn  \,
~~~~~~~~~~~~~~~~~~~~~~~~~~~~~~~~~~~~~~~~~~~~~~(2.10d) \\
&&~~~~B( \mu ) A( \lambda ) 
= \frac{\mu^2 q - \lambda^2 q^{-1}}{\mu^2 -\lambda^2 
- i \epsilon } A( \lambda ) B( \mu ) \,  , \nn 
~~~~~~~~~~~~~~~~~~~~~~~~~~~~~~~~~~~~~~~~~~~~~~~~~~(2.10e)\\
&&~~~~B( \mu ) B^\dagger( \lambda ) = \tau(\lambda,\mu) 
 B^\dagger( \lambda ) B( \mu ) + 4 \pi  \hbar  \lambda \mu \, \delta(
\lambda^2 - \mu^2 ) A^\dagger( \lambda ) A( \lambda ) \, , \nn  \,
~~~~~~~~~~~~~~  (2.10f)
\eea
\addtocounter{equation}{1}  
where $\tau( \lambda, \mu ) = \left[ \, 1 + \frac{8  \hbar^2 \xi^2 \lambda^2
\mu^2}{{( \lambda^2 - \mu^2 )}^2} - \frac{4 \hbar^2 \xi^2 \lambda^2
\mu^2}{( \lambda^2 - \mu^2 - i \epsilon ) ( \lambda^2 - \mu^2 + i \epsilon
)} \, \right]$.

Due to eqn.(2.10a) it follows that all operator 
valued coefficients occurring in the expansion of 
 $\ln A(\lambda)$ in powers of $\lambda $
must commute among themselves. Consequently,
 $\ln A(\lambda)$ may be treated as the generator of 
conserved quantities for the quantum integrable DNLS model.
For the purpose of diagonalising these quantum 
conserved quantities, we first 
 notice that the commutation relation 
(2.10f) contains product of singular functions $( \lambda^2
- \mu^2 - i \epsilon )^{-1}( \lambda^2 - \mu^2 + i \epsilon )^{-1} $, 
which does not make sense at the limit $\lambda 
\rightarrow \mu$.  As a result, actions of operators 
$B^\dagger(\lambda), B(\mu) $ are not well defined on the Hilbert space
 [4,25] and generate states which are 
not normalised on the $\delta$-function.
  However, it is well known that,
one can avoid this type of problem in the case of NLS
model by considering the quantum analogue of classical reflection
operators [3,26]. So, for the case of 
DNLS model also we consider a reflection operator given by 
\beq 
R^\dagger(\lambda ) ~=~ B^\dagger(\lambda) {( A^\dagger( \lambda 
))}^{-1}\, 
\label{b11}
\eeq
and its adjoint $R(\lambda)$. By using eqns.(2.10a-f), 
we find that such reflection operators satisfy well defined 
commutation relations like [21]
\bea
&&R^{\dagger}(\lambda)R^{\dagger}(\mu) = S^{-1}(\lambda , \mu) \, 
R^{\dagger}(\mu) R^{\dagger}(\lambda) \, , \nn \\
&&R(\lambda)R(\mu) 
= S^{-1}(\lambda , \mu) \, R(\mu) R(\lambda) \, , \nn \\
&&R^{\dagger}(\lambda)R(\mu) 
= S(\lambda , \mu) \, R(\mu) R^{\dagger}(\lambda) + 
4 \pi \hbar\lambda^2 \delta(\lambda^2 - \mu^2) \, ,
\label {b12}
\eea
where 
\beq
S(\lambda , \mu) \, = \,  { {\lambda^2 q - \mu^2 q^{-1}} \over 
{\lambda^2 q^{-1} - \mu^2 q }} \,. 
\label{b13}
\eeq
It is evident that these commutation relations are encoded in a 
form of Zamolodchikov-Faddeev algebra [3,27] and 
 $S(\lambda , \mu) $ (\ref {b13}) represents 
the nontrivial $S$-matrix element of two-body scattering between the 
related quasi-particles.  It is easy to check that this 
$S(\lambda , \mu) $ satisfies the relations 
\beq 
S^{-1}(\lambda , \mu) = S(\mu ,\lambda) = S^*(\lambda , \mu) \, ,
\label {b14}
\eeq
and remains nonsingular at the limit $\lambda \rightarrow \mu$.  
As a result,  the action of operators like 
$R^\dagger(\lambda) $ on the vacuum would produce well defined
states which can be normalised on the $\delta$-function.

The commutation relation between
$A(\lambda)$ and $R^{\dagger}(\mu)$ may be derived 
by using eqns.(2.10b) and (2.10d) as
\bea
~~~~A(\lambda)R^{\dagger}(\mu) = \frac{\mu^2q - \lambda^2q^{-1}}{\mu^2 -
\lambda^2 - i\epsilon}R^{\dagger}( \mu )A( \lambda ) \, . 
\label {b15}
\eea
 By applying the above commutation relation and also using 
 $ A(\lambda) \v 0 \r = \v 0 \r$, it can be shown that
\beq
 A(\lambda ) \, \v \mu_1 ,\mu_2  , \cdots , \mu_N  \r ~=~
\prod_{r=1}^N  \left( { \mu_r^2 q - \lambda^2 q^{-1} 
\over \mu_r^2  - \lambda^2 -i \epsilon } \right)  \, 
  \v \mu_1 ,\mu_2  , \cdots , \mu_N  \r \, ,
\label {b16}
\eeq
where 
  $\v \mu_1 ,\mu_2  , \cdots , \mu_N  \r \equiv 
 R^\dagger(\mu_1) R^\dagger(\mu_2) \cdots R^\dagger(\mu_N) \v 0 \r $ and 
$\mu_j$s are all distinct real or pure imaginary numbers.
Thus the states 
  $\v \mu_1 ,\mu_2  , \cdots , \mu_N  \r $ diagonalise the generator of 
conserved quantities for the quantum DNLS model.
 However, by using eqn.(\ref {b16}), one finds that  the 
eigenvalues corresponding to different expansion coefficients of 
 $ \ln A(\lambda) $ would be complex quantities in general. 
To make the eigenvalues real, we define another operator 
${\hat A}(\lambda)$ through the relation: 
${\hat A}(\lambda) \equiv 
  A(\lambda  e^{ -{i \alpha\over 2}}) $ and expand $\ln {\hat A}(\lambda)$ as
\beq
 \ln {\hat A}(\lambda) = 
\sum_{n=0}^{\infty} \frac{ i \,{\cal C}_n}{\lambda^{2n}}  \, .
\label {b17}
\eeq
With the help of eqns.(\ref {b16}) and (\ref {b17}), one can easily find out
the real eigenvalues associated with all ${\cal C}_n$s:
\bea
&& ~~~~~~~~~~~{\cal C}_0 \v \mu_1 ,\mu_2  , \cdots , \mu_N  \r ~=~\alpha N \,
  \v \mu_1 ,\mu_2  , \cdots , \mu_N  \r \, , \nn 
~~~~~~~~~~~~~~~~~~~~~~~~~~~~~~~~~~~(2.18a) \\
&& ~~~~~~~~~~~{\cal C}_n \v \mu_1 ,\mu_2  , \cdots , \mu_N  \r ~=~
\frac {2}{n} \sin(\alpha n) \Big \{ \sum_{j=1}^N \mu_j^{2n} \Big \}  \,
  \v \mu_1 ,\mu_2  , \cdots , \mu_N  \r \, , \nn 
\,  ~~~~~~~~~~~~~~(2.18b) 
\eea
\addtocounter{equation}{1} 
where $n \geq 1$.
Till now it is assumed that $\mu_j$s are some real or pure
imaginary parameters, for which 
 $ \v \mu_1 ,\mu_2  , \cdots , \mu_N  \r $ represents a scattering state.
We can also construct the quantum soliton states or bound states
 for DNLS model by choosing complex values of $\mu_j$ given by [19,21]
\beq
  \mu_j ~=~ \mu \, \exp \left[  i \alpha \left( {N+1 \over 2} -j \right)
\right] \, ,
\label {b19}
\eeq
where $\mu $ is a real or pure imaginary
 parameter and $j\in [1,2, \cdots N]$. Similar to 
the case of scattering states, one can find out
the real eigenvalues corresponding to all ${\cal C}_n$s for these
 quantum soliton states of DNLS model.

Thus, by applying QISM, it is possible to
 obtain the exact eigenvalues as well as
eigenstates for the quantum conserved quantities of DNLS model
which are defined formally through the expansion (\ref{b17}).
However, the important problem of expressing these conserved 
quantities through  basic field operators like $\psi (x)$ and
$\psi^\dagger (x)$ has not been explored so far.  
In analogy with the classical case, 
 ${\cal C}_0$, ${\cal C}_1$ and ${\cal C}_2$ should 
be related to the number operator, momentum operator and 
the Hamiltonian of the quantum DNLS model respectively.
So it should be particularly 
interesting to find out the explicit form of these
first three conserved quantities. To this end, we shall 
study quantum Jost solutions of the DNLS model.

\noindent \section 
{Jost Solutions of quantum DNLS model}
\renewcommand{\theequation}{3.{\arabic{equation}}}
\setcounter{equation}{0}

\medskip

It may be recalled that the differential equations 
satisfied by the Jost solutions of quantum NLS model 
are defined through the corresponding Lax operator [11].
As a result all Jost solutions of NLS model, 
defined through boundary conditions 
 at $x \rightarrow + \infty $ or $ x \rightarrow - \infty$,
 satisfy exactly the same form of coupled differential equations. 
At present, however,
we shall not directly use the Lax operator (\ref {a5}) for obtaining the 
differential equations associated with Jost solutions of quantum 
DNLS model.  Instead, we shall identify appropriate elements of the matrices 
${\cal T}_+( x,\lambda )$ (2.6a) and ${\cal T}_-( x,\lambda )$ (2.6b) as
 Jost solutions corresponding to boundary conditions 
 at $x \rightarrow + \infty $ and $ x \rightarrow - \infty$ respectively. 
The differential equations satisfied by 
${\cal T}_+( x,\lambda )$  and ${\cal T}_-( x,\lambda )$ will 
give us in a natural way the differential equations for 
 Jost solutions corresponding to boundary conditions 
 at $x \rightarrow + \infty $ and $ x \rightarrow - \infty$ respectively. 
It will turn out that, contrary to the case of NLS model,
 quantum Jost solutions 
of DNLS model associated with boundary conditions at $x\rightarrow
+\infty $ and $x\rightarrow - \infty$ satisfy different types of 
coupled differential equations. 
Due to eqn.(\ref {b5}), the elements of monodromy matrix (\ref {b7})
can be expressed as Wronskians of such Jost solutions. 

To proceed in the above mentioned way,
let us express ${\cal T}_-( x,\lambda )$ (2.6b) in elementwise form as 
\bea
{\cal T}_-( x,\lambda ) = \pmatrix {\phi_1( x, \lambda ) & \bar{\phi}_1( x,
\lambda ) \cr
 \phi_2( x, \lambda ) & \bar{\phi}_2( x, \lambda )} \, ,
\label{c1}
\eea
where 
${\phi}(x, \lambda) \equiv 
\pmatrix {{\phi}_1( x, \lambda ) \cr {\phi}_2( x, \lambda ) } $
and $\bar{\phi}(x, \lambda) \equiv 
\pmatrix {\bar{\phi}_1( x, \lambda ) \cr \bar{\phi}_2( x, \lambda ) } $
are two Jost solutions corresponding to boundary conditions at
  $ x \rightarrow - \infty$.  Due to eqn.(2.2a),
${\cal T}_-( x,\lambda )$ satisfies a differential equation given by
\bea
\partial_x{\cal T}_-( x, \lambda ) \, = \,
 : {\cal U}_q( x , \lambda ){\cal T}_-( x, \lambda ) : \, .
\label{c2}
\eea
Substituting the explicit form of ${\cal T}_-( x,\lambda )$ 
(\ref{c1}) to (\ref{c2}),  we find that the components of 
${\phi}(x, \lambda) $ and $\bar{\phi}( x, \lambda )$ satisfy 
exactly the same form of coupled differential equations given by
\bea
&&\partial_x\rho_1( x, \lambda ) \, = \, 
- \frac{i\lambda^2}{4}\rho_1( x, \lambda ) +
 if\psi^\dagger( x ) \rho_1( x, \lambda)\psi( x ) 
+ i\xi\lambda \psi^\dagger( x )\rho_2( x, \lambda ) \, , \nn \\
&&\partial_x\rho_2( x, \lambda ) \, = \,  
 \frac{i\lambda^2}{4}\rho_2( x, \lambda ) 
-ig\psi^\dagger( x ) \rho_2( x, \lambda)\psi( x ) 
+ i\lambda\rho_1( x,\lambda )\psi(x) \, ,
\label{c3}
\eea
where $\pmatrix {\rho_1( x, \lambda ) \cr
                         \rho_2( x, \lambda )} $ may be chosen either as
$ \pmatrix { \phi_1( x, \lambda ) \cr \phi_2( x, \lambda )} $ 
or as $ \pmatrix { \bar{\phi}_1( x, \lambda ) \cr
             \bar{\phi}_2( x, \lambda )} $.
Thus $\rho( x, \lambda ) \equiv \pmatrix {\rho_1( x, \lambda ) \cr
                         \rho_2( x, \lambda )} $ 
represents the general form of Jost solutions defined through boundary
conditions at $x \rightarrow {-\infty}$.  Next, 
by taking the $x\rightarrow - \infty$ limit of 
${\cal T}_-( x,\lambda )$ (2.6b), we obtain
\bea
{\cal T}_-( x, \lambda ) 
~ \stackrel{x \rightarrow \, - \infty}{\longrightarrow} ~
  e^{-{\frac{ i \lambda^2 x }{4}} \sigma_3 } \, .
\label{c4}
\eea
Substituting the matrix form of 
${\cal T}_-( x, \lambda )$ (\ref {c1}) to the relation (\ref {c4}),
we obtain the boundary conditions associated with Jost solutions 
${\phi}(x, \lambda)$ and $\bar{\phi}( x, \lambda )$ as
\beq
\pmatrix {\rho_1( x, \lambda ) \cr
           \rho_2 ( x, \lambda )} 
~ \stackrel{x \rightarrow \, - \infty}{\longrightarrow} ~
\pmatrix {\rho_1^0 e^{-\frac{i\lambda^2 x}{4}} \cr
          \rho_2^0 e^{\frac{i\lambda^2 x}{4}}} \, ,
\label{c5}
\eeq
where $\rho_1^0 = \one , \rho_2^0 = 0$ for $\rho( x, \lambda ) = \phi( x,
\lambda )$ and $\rho_1^0 = 0, \rho_2^0 = \one $ for $\rho( x, \lambda ) =
\bar{\phi}( x, \lambda )$.  Using the boundary conditions
(\ref{c5}), we can convert the differential equations (\ref{c3})
to their integral forms as 
\bea
&& \rho_1( x, \lambda ) \, = \, \rho_1^0e^{- \frac{i\lambda^2x}{4}} +
i\int_{-\infty}^x dz \, e^{\frac{i\lambda^2}{4}(z-x)}
\left \{ \, f\psi^\dagger(z)\rho_1(z, \lambda )\psi(z) 
+ \xi\lambda\psi^\dagger(z)\rho_2(z,\lambda) \,\right \} \, , \nn  \\
%&&~~~~~~~~~~~~~~~~~~~~~~~~~~~~~~~~~~~~~~~~~~~~~~~~~~~~~~~~~~
%~~~~~~~~~~~~~~~~~~~~~~~~~~~~~~~~~~~~~~~~~~~~ (3.6a)\nn \\
&& \rho_2( x, \lambda ) = \rho_2^0e^{\frac{i\lambda^2x}{4}} +
i\int_{-\infty}^x dz \, e^{\frac{i\lambda^2}{4}(x-z)}
\left \{ \, -g\psi^\dagger(z)\rho_2(z,\lambda)\psi(z) 
+\lambda\rho_1(z, \lambda )\psi(z) \,\right \} \, . \nn  \\
&&~~~~~~~~~~~~~~~~~~~~~~~~~~~~~~~~~~~~~~~~~~~~~~~~~~~~~~~~~~~~~~~~
~~~~~~~~~~~~~~~~~~~~~~~~~~~~~~~~~~~~~(3.6a,b) \nn
\eea
\addtocounter{equation}{1}
With the help of these integral relations it is easy to show that, 
for the case of real $\lambda$, the components of Jost
solutions $\phi(x,\lambda)$ and $\bar{\phi}(x,\lambda)$ are related
as
\bea
\bar{\phi}_1( x, \lambda ) = -\xi\phi_2^\dagger( x, \lambda ) \, ,
~~\bar{\phi}_2( x, \lambda ) = \phi_1^\dagger( x, \lambda ) \, .
\label{c7}
\eea

Next we try to find out the differential equations for the
 Jost solutions corresponding to boundary conditions 
 at $x \rightarrow + \infty $. To this end,
 we express ${\cal T}_+( x,\lambda )$ (2.6a) in elementwise form as 
\bea
{\cal T}_+( x,\lambda ) = \pmatrix {\chi_2( x, \lambda ) &  - \chi_1(x,
\lambda) \cr \bar {\chi}_2(x, \lambda) &  - \bar{\chi}_1(x,\lambda)} \, ,
\label{c8}
\eea
where ${\chi}(x, \lambda) \equiv 
\pmatrix {{\chi}_1( x, \lambda ) \cr {\chi}_2( x, \lambda ) } $
and $\bar{\chi}(x, \lambda) \equiv 
\pmatrix {\bar{\chi}_1(x,\lambda) \cr \bar{\chi}_2( x, \lambda ) } $
represent two Jost solutions corresponding to boundary conditions at
  $ x \rightarrow + \infty$.  Due to relation (2.2b),
${\cal T}_+( x,\lambda )$ satisfies a differential equation given by
\bea
\partial_x{\cal T}_+( x, \lambda ) \, = \, -
:{\cal T}_+( x, \lambda ) {\cal U}_q( x , \lambda ):
\, .
\label{c9}
\eea
Substituting the elementwise form of ${\cal T}_+( x,\lambda )$ 
(\ref{c8}) to (\ref{c9}),  it is easy to see that the components of 
${\chi}(x, \lambda) $ and $\bar{\chi}( x, \lambda )$ satisfy 
exactly the same form of coupled differential equations given by
\bea
&&\partial_x\tau_1( x, \lambda ) \, = \,  
- \frac{i\lambda^2}{4}\tau_1( x, \lambda ) + ig \, \psi^\dagger( x )
\tau_1( x, \lambda) \psi(x) + i\xi\lambda \,
\psi^\dagger(x)\tau_2( x, \lambda ) \, , \nn \\
&&\partial_x\tau_2( x, \lambda ) \, = \,
 \frac{i\lambda^2}{4}\tau_2( x, \lambda ) - if \, \psi^\dagger( x )
\tau_2( x, \lambda) \psi(x) + i\lambda \,
\tau_1( x, \lambda ) \psi(x) \, , \nn \\
\label{c10}
\eea
where $\pmatrix {\tau_1( x, \lambda ) \cr
                         \tau_2( x, \lambda )} $ may be chosen as either
$ \pmatrix { \chi_1( x, \lambda ) \cr \chi_2( x, \lambda )} $ 
or  $ \pmatrix { \bar{\chi}_1( x, \lambda ) \cr
             \bar{\chi}_2( x, \lambda )} $.
Thus $\tau (x,\lambda) \equiv \pmatrix {\tau_1( x, \lambda ) \cr
                         \tau_2( x, \lambda )} $ 
represents the general form of Jost solutions defined through boundary
conditions at $x \rightarrow {+\infty}$.  Next, 
by taking the $x\rightarrow + \infty$ limit of 
${\cal T}_+( x,\lambda )$ (2.6a), we obtain
\bea
{\cal T}_+( x, \lambda ) 
~ \stackrel{x \rightarrow \,  + \infty}{\longrightarrow} ~
e^{ {\frac{ i \lambda^2 x }{4}} \sigma_3 } \, .
\label{c11}
\eea
Substituting the explicit form of 
${\cal T}_+( x, \lambda )$  (\ref {c8}) to the above relation,
it is easy to find out the 
boundary conditions associated with Jost solutions 
${\chi}(x,\lambda)$ and $\bar{\chi}(x,\lambda)$ as
\beq
\pmatrix {\tau_1( x, \lambda ) \cr
           \tau_2 ( x, \lambda )} 
~ \stackrel{x \rightarrow \,  + \infty}{\longrightarrow} ~
\pmatrix {\tau_1^0 e^{-\frac{i\lambda^2 x}{4}} \cr
          \tau_2^0 e^{\frac{i\lambda^2 x}{4}}} \, ,
\label{c12}
\eeq
where $\tau_1^0 = 0, \, \tau_2^0 = \one $ for $\tau(x,\lambda) = \chi( x,
\lambda )$ and $\tau_1^0 = - \one , \, \tau_2^0 = 0$ for $\tau(x,\lambda) =
\bar{\chi}(x,\lambda)$.  Using the boundary conditions
(\ref{c12}), we can convert the differential equations (\ref{c10})
to their integral forms as 
\bea
&& \tau_1(x,\lambda) \, = \, \tau_1^0e^{-\frac{i\lambda^2x}{4}} -
i\int_x^{\infty} dz \, e^{\frac{i\lambda^2}{4}(z-x)}
\left \{ \, g \, \psi^\dagger(z)\tau_1(z, \lambda )\psi(z) 
+ \xi\lambda \, \psi^\dagger(z)\tau_2(z,\lambda ) \,\right \} \, , \nn  \\
&& \tau_2( x, \lambda ) = \tau_2^0e^{\frac{i\lambda^2x}{4}} -
i\int_x^{\infty} dz \, e^{\frac{i\lambda^2}{4}(x-z)}
\left \{ \, -f \, \psi^\dagger( z
)\tau_2(z, \lambda)\psi(z) + \lambda \, \tau_1(z, \lambda )
\psi(z) \,\right \} \, . \nn  \\
&&~~~~~~~~~~~~~~~~~~~~~~~~~~~~~~~~~~~~~~~~~~~~~~~~~~~~~~~~~~~~~~~~
~~~~~~~~~~~~~~~~~~~~~~~~~~~~~~~~~~~~(3.13a,b) \nn
\eea
\addtocounter{equation}{1}
By using these integral relations it is easy to show that, 
for the case of real $\lambda$, the components of Jost
solutions $\chi( x, \lambda )$ and $\bar{\chi}( x, \lambda )$ are related
as
\bea
\bar{\chi}_1( x, \lambda ) = -\chi_2^\dagger( x, \lambda ) \, ,
~~\bar{\chi}_2( x, \lambda ) = {1\over \xi} \chi_1^\dagger( x, \lambda ) \, .
\label{c14}
\eea

Comparing eqns.(\ref {c10}) and (\ref {c3}), 
 we notice that quantum Jost solutions 
of DNLS model, associated with boundary conditions at $x\rightarrow
+\infty $ and $x\rightarrow - \infty$, satisfy two different sets of 
coupled differential equations. These two sets of differential 
equations are related to each other through an interchange of $f$ and $g$.
However, since both $f$ and $g$ coincide with the 
coupling constant $\xi$ at $\hbar \rightarrow 0$ limit, 
 eqns.(\ref {c3}) and (\ref {c10}) have an identical form
at this classical limit.  It may also be observed that, 
due to vanishing boundary condition
on the basic field variables, 
eqns. (\ref {c3}) and (\ref {c10}) have the same asymptotic form 
at $\v x \v \rightarrow \infty $ limit. 

Now we want to express the elements of quantum monodromy matrix (\ref{b7})
in terms of Jost solutions as obtained above. To this end, we substitute 
the elementwise form of ${\cal T}_-( x,\lambda )$ (\ref{c1}) and 
${\cal T}_+( x,\lambda )$ (\ref{c8}) to eqn.(\ref{b5}) and 
compare it with (\ref{b7}). In this way, we obtain
\bea
~~~~~~~~~~~~~~~~~~&&A( \lambda ) = \chi_2( x, \lambda )
\phi_1( x, \lambda ) - \chi_1( x, \lambda)\phi_2(x, \lambda), \nn \\
&&A^\dagger(\lambda) = \bar{\chi}_2(x,\lambda)\bar{\phi}_1(x, \lambda)
 - \bar{\chi}_1( x, \lambda )\bar{\phi}_2(x, \lambda ), \nn \\
~~~~~~~~~~~~~~~~~~&&B( \lambda ) = \bar{\chi}_2( x, \lambda )
\phi_1( x, \lambda ) -
\bar{\chi}_1(x, \lambda )\phi_2( x, \lambda ),\nn \\
&&B^\dagger( \lambda ) =
-\frac{1}{\xi}\chi_2( x, \lambda )\bar{\phi}_1( x, \lambda ) +
\frac{1}{\xi}\chi_1( x, \lambda )\bar{\phi}_2( x, \lambda ).
~~~~~~~~~(3.15 a,b,c,d) \nn
\eea
\addtocounter{equation}{1}  
Since the l.h.s. of eqns.(3.15a-d) 
do not depend at all on the variable $x$,
the r.h.s. of these equations should also be independent of this
 variable (in spite of its explicit appearance).  By taking 
 $x\rightarrow +\infty $ or $x\rightarrow - \infty$ limit  in the 
r.h.s. of eqns.(3.15a-d) and using boundary conditions (\ref {c5})
or (\ref {c12}) respectively, we obtain
\bea
~~~~~~~~~~~~~~~~&&A( \lambda ) = \lim_{x\rightarrow -\infty}e^
{-\frac{i\lambda^2}{4}x}\chi_2(
x, \lambda ) = \lim_{x\rightarrow +\infty}e^{\frac{i\lambda^2}{4}x}
\phi_1(x, \lambda ) \, ,\nn ~~~~~~~~~~~~~~~~~~~(3.16a) \\
~~~~~~~~~~~~~~~~&&A^\dagger( \lambda ) = -\lim_{x\rightarrow -\infty}e^
{\frac{i\lambda^2}{4}x}
\bar{\chi}_1( x, \lambda ) = \lim_{x\rightarrow +\infty}e^{-\frac{i\lambda^2}
{4}x}\bar{\phi}_2( x, \lambda ) \, ,\nn ~~~~~~~~~~~~~~~(3.16b) \\
~~~~~~~~~~~~~~~~&&B( \lambda ) = \lim_{x\rightarrow -\infty}e^
{-\frac{i\lambda^2}{4}x}
\bar{\chi}_2( x, \lambda ) =  \lim_{x\rightarrow +\infty}e^{-\frac{i\lambda^2}
{4}x}\phi_2( x, \lambda )\, ,\nn ~~~~~~~~~~~~~~~~~(3.16c) \\
~~~~~~~~~~~~~~~~&&B^\dagger( \lambda ) = \frac{1}{\xi} \lim_{x\rightarrow -
\infty}e^{\frac
{i\lambda^2}{4}x}
\chi_1( x, \lambda ) = -\frac{1}{\xi} \lim_{x\rightarrow +\infty}e^{\frac
{i\lambda^2}
{4}x}\bar{\phi}_1( x, \lambda) \, .\nn ~~~~~~~~~~~~( 3.16d) 
\eea
\addtocounter{equation}{1}

Next, let us define the quantum Wronskian 
associated with the general form of Jost solutions $\tau (x,\lambda)$ and 
 $\rho (x,\lambda)$ as 
\bea
\Lambda_{\rho , \tau }(x,\lambda) = 
\tau_2( x, \lambda) \rho_1( x,\lambda)
- \tau_1(x,\lambda) \rho_2(x,\lambda) \, .
\label{c17}
\eea
Comparing eqns.(3.15) and (\ref {c17}) for all possible choice of 
$\tau (x,\lambda)$ and $\rho (x,\lambda)$,
we find that 
\bea
&&A(\lambda) = \Lambda_{\phi ,\chi}(x,\lambda) \, , ~~~~
B(\lambda) = \Lambda_{\phi , {\bar \chi}}(x,\lambda) \, , \nn \\
&&A^\dagger(\lambda) = \Lambda_{\bar{\phi} ,{\bar \chi}}(x,\lambda)  \, , 
~~~~ B^\dagger(\lambda) = - {1\over \xi} \, 
\Lambda_{{\bar \phi}, \chi}(x,\lambda) \, . 
\label{c18}
\eea
Thus the quantum Wronskian (\ref {c17}) represents all
elements of the monodromy matrix (\ref {b7}) in a general form.

Since the elements of monodromy matrix (\ref {b7}) do not depend 
on the variable $x$, the quantum Wronskian (\ref {c17}) must also be
independent of this variable.  However, we should be able to demonstrate
 this fact in a direct way by showing that 
$\Lambda_{\rho , \tau }(x,\lambda)$  has a vanishing derivative 
with respect to the variable $x$. To this end, 
we consider a general type of quantum integrable field model 
whose Jost solutions $\rho (x, \lambda)$ and 
$\tau(x, \lambda)$ satisfy differential equations given by
\beq
\d_x \rho(x, \lambda) \, = \, 
 :{\cal L}^{-}(x, \lambda) \rho(x, \lambda) : \, ,
~~\d_x \tau (x, \lambda) \, = \,
 :{\cal L}^{+}(x, \lambda) \tau (x, \lambda): \, ,
\label {c19}
\eeq
${\cal L}^{\pm}(x, \lambda)$ being some $(2\times 2)$-matrices with elements 
${\cal L}_{ij}^{\pm}(x, \lambda)$. As before, 
the quantum Wronskian associated with this general case may be defined
 through eqn.(\ref {c17}).  For the sake of convenience,
let us ignore at present the effect of normal ordering in eqn.(\ref {c19}) 
and treat all quantum variables as commuting classical variables. 
In this way it can be easily shown that, 
 the derivative of Wronskian (\ref {c17})
with respect to the variable $x$ will vanish if the elements of 
${\cal L}^{+}(x, \lambda)$ and ${\cal L}^{-}(x, \lambda)$ are related as 
\beq
{\cal L}^{-}_{11}(x,\lambda) = - {\cal L}^{+}_{22}(x, \lambda) \, , ~~
{\cal L}^{-}_{22}(x,\lambda) = - {\cal L}^{+}_{11}(x, \lambda) \, , ~~
{\cal L}^{-}_{ij}(x,\lambda) =  {\cal L}^{+}_{ij}(x, \lambda) \, ,
\label {c20}
\eeq
where $i\neq j$. Thus it follows that,
${\cal L}^{+}(x, \lambda)$ would coincide with 
${\cal L}^{-}(x, \lambda)$ when it satisfies the
traceless condition. Since the Lax operators of quantum NLS model and 
almost all other integrable systems satisfy this traceless condition,
${\cal L}^{+}(x, \lambda)$ and 
${\cal L}^{-}(x, \lambda)$ coincide for these cases. However, the 
quantum Lax operator (\ref {a5}) 
of DNLS model does not satisfy this condition. Consequently,
the corresponding ${\cal L}^{+}(x, \lambda)$ and 
${\cal L}^{-}(x, \lambda)$ matrices 
should not coincide with each other. Expressing eqns.(\ref {c3}) 
and (\ref {c10}) in matrix form, we find that 
${\cal L}^{-}(x, \lambda)$ matrix of DNLS model is same as 
${\cal U}_q( x , \lambda )$ (\ref {a5}) and 
${\cal L}^{+}(x, \lambda)$ may be obtained from 
${\cal U}_q( x , \lambda )$ by interchanging $f$ and $g$. 
Since these matrices 
satisfy the relation (\ref {c20}), we may conclude that the 
Wronskian (\ref {c17}) of DNLS model 
has a vanishing derivative with respect to the variable $x$.
 A more rigorous proof about the 
coordinate independence of this Wronskian, 
taking into account the noncommutative nature of quantum operators, 
will be given in Sec.5 of this article. 

\vspace{1cm}
\noindent \section {Spectrum Generating Algebra for DNLS Model }
\renewcommand{\theequation}{4.{\arabic{equation}}}
\setcounter{equation}{0}

\medskip

By following the approach of Ref.11, here we shall propose the 
`fundamental relation' for the DNLS model and explore its 
connection with the spectrum generating algebra.
In analogy with the quantum Wronskian (\ref {c17}),
let us define another operator associated with the Jost solutions 
of DNLS model as 
\bea
\Gamma_{\rho,\tau}( x,\lambda )  \, =  \,
\tau_2( x, \lambda ) \rho_1( x, \lambda )
+ \tau_1( x, \lambda ) \rho_2( x, \lambda )
  \, .
\label{d1}
\eea
This $\Gamma_{\rho,\tau}(x,\lambda) $ and 
 quantum Wronskian (\ref {c17}) are two basic ingredients which are needed
for defining the fundamental relation of DNLS model.
Now we propose that,
the quantum conserved quantities ($I_n$) of DNLS model 
would annihilate the vacuum state and obey the 
fundamental relation given by
\bea
\Big[ I_n , \Lambda_{\rho,\tau}( \lambda ) \Big] 
 &=& \frac{\hbar\lambda^{2n}}{2^{n+1}} \int_{-\infty}^{+\infty} 
\d_y \Gamma_{\rho,\tau}(y,\lambda)  \,   dy  \nn \\
&=& \frac{\hbar\lambda^{2n}}{2^{n+1}} \, 
\Big \{ \, \Gamma_{\rho,\tau}( +\infty, \lambda ) 
- \Gamma_{\rho,\tau}(-\infty, \lambda ) \, \Big \} \, ,
\label{d2}
\eea
where $n$ is any nonnegative integer.
Since $\Lambda_{\rho,\tau}(x,\lambda)$ (\ref {c17})
 does not depend on the coordinate $x$,  we have suppressed 
this variable in the l.h.s. of above relation.  

Next,  we shall discuss how the 
fundamental relation (\ref {d2}) leads to 
 the spectrum generating algebra for all quantum conserved quantities of 
DNLS model. To this end, it is needed to find out the 
$x \rightarrow \pm \infty$ limit of 
$ \Gamma_{\rho,\tau}( x,\lambda ) $.
For all possible choices of $\rho$ and $\tau$, 
$\Gamma_{\rho,\tau}(x,\lambda)$ (\ref {d1}) may be explicitly 
written as 
\bea
~~~~~~~~~~~~~~~~~~~~~~~&&\Gamma_{\phi,\chi}( x,\lambda ) = 
 \, \chi_2( x, \lambda )\phi_1( x, \lambda )
+ \chi_1( x, \lambda )\phi_2( x, \lambda ) \, ,\nn ~~~~~~~~~~~~~~~
~ \, (4.3a )\\
~~~~~~~~~~~~~~~&&\Gamma_{\bar{\phi},\bar{\chi}}( x,\lambda ) =
\,  \bar{\chi}_2( x, \lambda )
\bar{\phi}_1( x, \lambda ) + \bar{\chi}_1( x, \lambda )\bar{\phi}_2
(x,\lambda) \,  ,\nn~~~~~~~~~~~~~~~~ \, ( 4.3b ) \\
~~~~~~~~&&\Gamma_{\phi,\bar{\chi}}( x,\lambda ) =
 \, \bar{\chi}_2( x, \lambda )
\phi_1( x, \lambda ) + \bar{\chi}_1( x, \lambda )\phi_2( x, \lambda ) 
\, ,
\nn ~~~~~~~~~~~~~~~~ \, (4.3c) \\
~~~~~~~~~~~~&&\Gamma_{\bar{\phi},\chi}( x,\lambda ) = 
 \, \chi_2( x, \lambda )
\bar{\phi}_1( x, \lambda ) + \chi_1( x, \lambda )\bar{\phi}_2( x, \lambda
) \, .
\nn ~~~~~~~~~~~~~~~~~(4.3d)
\eea
\addtocounter{equation}{1}
Substituting the asymptotic forms 
of Jost solutions (\ref{c5}), (\ref{c12}) to the 
$x \rightarrow \pm \infty$ limits of relations (4.3a-d)
 and subsequently using (3.16a-d), we find that
\bea
&& \Gamma_{\phi,\chi}( \pm \infty, \lambda ) \, = \,  A( \lambda ) \, , ~~
\Gamma_{\bar{\phi},\bar{\chi}} ( \pm \infty, \lambda ) \, = \, 
- A^\dagger( \lambda ) \, ,  \nn \\
&& \Gamma_{\phi,\bar{\chi}}( \pm \infty, \lambda ) 
\, = \,  \mp B( \lambda ) \, , ~~
\Gamma_{\bar{\phi},\chi}( \pm \infty, \lambda ) \, = \,  \mp \xi
B^\dagger(\lambda) \, .
\label {d4}
\eea
Inserting (\ref{d4}) to the  
fundamental relation (\ref {d2}) and also using (\ref {c18}), we get 
\bea
~~~~~~~~~~~~~~~~~~ &&\Big[ I_n, A( \lambda ) \Big] = 0  \, , ~~~
\left[ I_n , A^\dagger( \lambda ) \right] = 0 , \nn
~~~~~~~~~~~~~~~~~~~~~~~~~~~~~~~~~~~~~~(4.5a,b) \\
~~~~~~~~~~~~~~~~~~&&\Big[ I_n, B( \lambda )\Big] = -
\frac{\hbar\lambda^{2n}}{2^n} B(\lambda) ,
~~~\left[I_n , B^\dagger(\lambda) \right] 
= \frac{\hbar\lambda^{2n}}{2^n} B^\dagger
( \lambda ).\nn
~~~~~~~~~~~~(4.5c,d )
\eea
\addtocounter{equation}{1}
 With the help of eqns.(4.5b,d), we can find out the commutation relation
between the quantum conserved quantities and 
reflection operators (\ref {b11}) as 
\bea
\left[ I_n , R^\dagger(\lambda) \right] \, = \,
 \frac{\hbar\lambda^{2n}}{2^n} \,  R^\dagger( \lambda ) \, .
\label{d6}
\eea
By using the above commutation relation and assuming that
 $I_n$s annihilate the vacuum state, it is easy to show that these
conserved quantities satisfy eigenvalue equations given by 
\bea
I_n \, \v \mu_1,\mu_2,\cdots ,\mu_N \r \, = \,
 \Big( \, \frac{\hbar}{2^n}\sum_{j=1}^N \mu_j^{2n} \, \Big) \v
 \,  \mu_1,\mu_2,\cdots ,\mu_N \r \, ,
\label{d7}
\eea
where $\v \mu_1,\mu_2,\cdots ,\mu_N \r \equiv R^{\dagger}(\mu_1)
R^{\dagger}(\mu_2)\cdots R^{\dagger}(\mu_N)\v 0 \r $.
Consequently,
the commutation relation (\ref {d6}) may be treated as the spectrum
generating algebra for the quantum conserved quantities of DNLS model.

It should be noted that, eigenstates of $I_n$ 
  are same as Bethe states which 
 we have already used in the 
framework of QISM to diagonalise the quantum conserved quantities 
  appearing in the expansion (\ref {b17}).
Thus, it is natural to expect a connection between these $I_n$s  
 and the conserved quantities 
 which are formally defined through the expansion (\ref {b17}). For
 establishing this connection,
let us assume that the Bethe states 
 $\v \mu_1,\mu_2,\cdots ,\mu_N \r $ represent a complete set of 
states in the corresponding Hilbert space. Thus two operators 
would coincide if they can be simultaneously 
diagonalised through these complete set of states and 
their eigenvalues always match with each other.
Comparing eqns.(\ref {d7}) with (2.18a,b), it is easy to find that
\bea
C_0 \, = \, \frac{\alpha}{\hbar} \, I_0 \, ,~~~ 
C_n \, = \,  \frac{2^{n+1}}{n\hbar} \sin(\alpha n) \, I_n \, .
\label{d8}
\eea
Substituting (\ref{d8}) to (\ref {b17}), we obtain 
the expansion of $\ln \hat{A}( \lambda )$ in terms of
$I_n$'s as
\bea
\ln \hat{A}( \lambda ) \, = \,  \frac{i\alpha}{\hbar} I_0  \, + \, 
\frac{i}{\hbar}\sum_{n=1}^\infty \, 
\frac{2^{n+1}}{n\lambda^{2n}}\sin(\alpha n) \, I_n \, .
\label{d9}
\eea

We can also define 
 conserved quantities for DNLS model
 through reflection operators as
\bea
I'_n = \frac{1}{2^{n+1}\pi}\int_0^\infty \mu^{2n-1}R^\dagger( \mu )
R( \mu ) d\mu.
\label{d10}
\eea
By using the commutation relations between reflection operators 
(\ref{b12}), which are derived in the framework of QISM, we obtain 
\bea
~~~~~~~~~~~~~~~~~~~~~~~~~~[ I'_n, I'_m ] \, = \,  0, 
~~~ [ I'_n, R^\dagger( \lambda ) ] \, = \, \frac
{\hbar \lambda^{2n}}{2^n} \,  R^\dagger( \lambda ) \, .\nn
~~~~~~~~~~~~~~~~~~~~~~~~(4.11a,b)
\eea
\addtocounter{equation}{1}
With the help of (4.11b), one can easily 
show that $\v \mu_1,\mu_2,\cdots ,\mu_N \r $ are eigenfunctions of
$I'_n$ with exactly the same eigenvalues as found in the
case of $I_n$ and conclude that $I_n = I'_n$.
Consequently, equation (\ref {d10}) yields 
an expression of $I_n$ through the reflection operators of DNLS model.

Finally, let us investigate whether
the fundamental relation may also lead to the spectrum of 
a general quantum integrable field model
whose Jost solutions satisfy the relations (\ref {c19}). For this 
purpose, we assume that
${\cal L}^{\pm}(x, \lambda)$ matrices have the following asymptotic
form  at $\v x \v \rightarrow \infty$ limit:
\bea
{\cal L}^{\pm}(x, \lambda) \longrightarrow i
 \pmatrix { l(\lambda) &  0 \cr 0 & - l(\lambda) } \, ,
\label{d12}
\eea
where $l(\lambda)$ is a function of the spectral parameter. Due to these  
 asymptotic forms of ${\cal L}^{\pm}(x, \lambda)$, the corresponding Jost
solutions can be defined through boundary conditions given by 
\beq
\rho (x,\lambda)
~ \stackrel{x \rightarrow \, - \infty}{\longrightarrow} ~
\pmatrix {\rho_1^0 e^{i l(\lambda)x} \cr
          \rho_2^0 e^{- i l(\lambda)x}} \, , ~~~~
\tau (x,\lambda)
~ \stackrel{x \rightarrow \, + \infty}{\longrightarrow} ~
\pmatrix {\tau_1^0 e^{i l(\lambda)x} \cr
          \tau_2^0 e^{- i l(\lambda)x}} \, .
\label{d13}
\eeq
Similar to the case of DNLS model, here we choose
 $\rho_1^0 = \one , \rho_2^0 = 0$ when $\rho( x, \lambda ) \equiv \phi( x,
\lambda )$,  $\rho_1^0 = 0, \rho_2^0 = \one $ when $\rho( x, \lambda ) 
\equiv \bar{\phi}( x, \lambda )$, 
 $\tau_1^0 = 0, \, \tau_2^0 = \one $ when $\tau(x,\lambda) \equiv \chi( x,
\lambda )$ and $\tau_1^0 = - \one , \, \tau_2^0 = 0$ when 
$\tau(x,\lambda) \equiv \bar{\chi}(x,\lambda)$. 
The quantum Wronskian and 
$\Gamma_{\rho,\tau}(x,\lambda)$ operator associated with these Jost
solutions are defined 
through eqns.(\ref {c17}) and (\ref {d1}) respectively. 
By treating quantum operators as commuting classical variables
and using the condition (\ref {c20}), 
we have already shown that 
 $\Lambda_{\rho,\tau}(x,\lambda)$ is 
 independent of the variable $x$. Here we assume that this 
 Wronskian would remain independent of $x$, even if the noncommuting 
nature of quantum operators are taken into account. 
Now we propose that hermitian 
conserved quantities (${\Im}_n$) associated with this general integrable 
field model satisfy fundamental relation of the form 
\bea
\Big[  {\Im}_n , \Lambda_{\rho,\tau}( \lambda ) \Big] 
= q_n(\lambda)
\Big \{ \, \Gamma_{\rho,\tau}( +\infty, \lambda ) 
- \Gamma_{\rho,\tau}(-\infty, \lambda ) \, \Big \} \, ,
\label{d14}
\eea
where $n$ is any nonnegative integer and $q_n(\lambda)$ is some real 
function of $\lambda$ whose explicit form 
 depends on the system concerned.  Taking 
 $x \rightarrow \pm \infty $ limits of $ \Lambda_{\rho,\tau}(x, \lambda )$ 
  (\ref {c17}) and using (\ref {d13}), we find that 
\bea
&&\Lambda_{\phi ,\chi}(\lambda) = \lim_{x\rightarrow -\infty}
e^{i l(\lambda)x} \chi_2(x, \lambda) = \lim_{x\rightarrow +\infty}
e^{- i l(\lambda)x} \phi_1(x, \lambda) \, , \nn \\
&&\Lambda_{{\bar \phi} ,\chi}(\lambda) = - \lim_{x\rightarrow -\infty}
e^{- i l(\lambda)x} \chi_1(x, \lambda) = \lim_{x\rightarrow +\infty}
e^{- i l(\lambda)x} {\bar \phi}_1(x, \lambda) \, .
\label {d15}
\eea
Similarly, by taking 
 $x \rightarrow \pm \infty $ limits of $ \Gamma_{\rho,\tau}(x, \lambda )$ 
  (\ref {d1}) and comparing them with (\ref {d15}),
it is easy to show that 
\beq
 \Gamma_{\phi ,\chi}(\pm \infty , \lambda ) \, = \, 
 % \Gamma_{\phi ,\chi}(- \infty , \lambda ) = 
 \Lambda_{\phi ,\chi}(\lambda)  \, , ~~~~
 \Gamma_{{\bar \phi} ,\chi}(\pm \infty , \lambda ) \, = \, \pm \,
 % - \Gamma_{{\bar \phi} ,\chi}(- \infty , \lambda ) = 
 \Lambda_{{\bar \phi} ,\chi}(\lambda)  \, .
\label {d16}
\eeq
Inserting (\ref {d16}) to (\ref {d14}), we find that 
\bea
~~~~~~~~~~~~~~~~~~
\Big [ \, {\Im}_n, \Lambda_{\phi ,\chi}(\lambda) \, \Big] \, = \,  0 \, , 
~~~~ \Big[ \, {\Im}_n, \Lambda_{{\bar \phi} ,\chi}(\lambda) \, \Big] \, =\, 
 2 q_n(\lambda) \, \Lambda_{{\bar \phi} ,\chi}(\lambda) \, .  \nn
~~~~~~~~~~~~~(4.17a,b)
\eea
\addtocounter{equation}{1}
By using (4.17b) and assuming that
 ${\Im}_n$s annihilate the vacuum state, we obtain the spectra for 
these conserved quantities as 
\bea
{\Im}_n \, \v \mu_1,\mu_2,\cdots ,\mu_N \r \, = \,
\left( 2 \sum_{i=1}^N q_n(\mu_i) \right)
 \,  \v \mu_1,\mu_2,\cdots ,\mu_N \r \, ,
\label{d18}
\eea
where $\v \mu_1,\mu_2,\cdots ,\mu_N \r \equiv 
 \Lambda_{{\bar \phi} ,\chi} (\mu_1)
 \Lambda_{{\bar \phi} ,\chi} (\mu_2) \cdots 
 \Lambda_{{\bar \phi} ,\chi} (\mu_N) \v 0 \r $. 
%Thus, equation (4.17b) represents the spectrum generating algebra 
%for all ${\Im}_n$.  
%Assuming that $\v \mu_1,\mu_2,\cdots ,\mu_N \r s$ form  a
%complete set of states, one may further conclude that all 
%${\Im}_n$s are commuting among themselves. 
%Therefore, ${\Im}_n$s may be treated as conserved quantities
%of a quantum integrable system. 

Thus, the fundamental relation (\ref {d14}) is powerful enough to 
  generate the spectra of conserved quantities for a class of 
  quantum integrable field models associated with Lax equations (\ref {c19}).
In the rest of this article, however, we shall restrict our attention 
only to quantum DNLS model and try to 
explicitly construct first few quantum conserved quantities which would 
satisfy the corresponding
fundamental relation (\ref {d2}). Necessary tools for such 
construction will be discussed in the next section.

\vspace{1cm}

\noindent \section { Commutation relations between the quantum
\hfil \break
 Wronskian and basic field operators}
\renewcommand{\theequation}{5.{\arabic{equation}}}
\setcounter{equation}{0}

\medskip

Since the quantum Wronskian (\ref {c17}) of DNLS model
is expressed as a bilinear 
function of Jost solutions, at first we consider 
 the commutation relations between these Jost solutions 
and basic field operators of the system. 
In analogy with the case of NLS model [11], one may take the arguments of 
Jost solutions and field operators 
at exactly the same space point and try to evaluate 
their commutation relations (e.g., commutators of the form
$[\rho_i(x,\lambda ),\psi(x)] \, $). 
By using the integral relations 
(3.6) and canonical commutation relations (\ref {a4}),
it can be easily checked that the commutators
$[\rho_i(x,\lambda ),\psi(x)]$  lead to indeterminant integrals of the form
 $\int_{-\infty}^x \delta(x-z) F(z) dz $, where $F(z)$ is some function
of $z$. Such indeterminant integrals, which also appear in the case 
of NLS model, may be fixed through a convention given by
 $\int_{-\infty}^x \delta(x-z) F(z) dz = {1\over 2} F(x)$ [11]. 
However, as will be explained shortly, the above mentioned
convention of fixing indeterminant integrals would lead to the violation 
of Jacobi identity in the case of DNLS model. So, instead of trying 
to calculate commutators of the form 
$[ \rho_i(x,\lambda ), \psi(x) ]$, at present we shall study 
commutators like $[ \rho_i(y,\lambda ), \psi(x) ]$ 
in the limit $y \rightarrow x$.

 To begin with, let us consider the commutators 
$\big[ \rho_i( y, \lambda ), \psi(x) \big] $ and 
$\big[ \rho_i( y, \lambda ),\psi^\dagger(x) \big] $
 in the region $y<x$.
For this case, all
   fields $\psi(z), \, \psi^\dagger(z)$ appearing in 
the integral relations (3.6a,b)
 would commute with $\psi(x), \, \psi^\dagger(x)$. Consequently, we obtain
 $\Big[ \rho_i( y, \lambda ), \psi( x ) \Big] = \left[ 
\rho_i( y, \lambda ), \psi^\dagger(x) \right] = 0$ in the region
 $y < x $. The $y\rightarrow x$ limit of these commutation 
relations may be expressed in the form 
\bea
\Big[ \rho_{i}( x'', \lambda ), \psi( x ) \Big] 
= \left[ \rho_{i}( x'', \lambda ),
\psi^\dagger( x ) \right] = 0 \, ,
\label{e1}
\eea
where the notation 
 $\rho_{i}( x'', \lambda ) \equiv \lim_{\epsilon\rightarrow 0+}
\rho_i( x - \epsilon , \lambda )$ is introduced and 
 $\epsilon\rightarrow 0+ $ limit is taken 
 \emph {after} evaluating all commutators.

Next, we consider  the commutators 
 $\big[ \rho_i( y, \lambda ), \psi( x ) \big] $
and $\big[ \rho_i( y, \lambda ), \psi^\dagger(x) \big] $
 in the region $y>x$.
 For this case, however, 
eqns.(3.6a,b) lead to rather complicated integral relations 
 which are difficult to solve 
in a closed form for arbitrary values of $x$ and $y$. 
So, for the sake of convenience,
we shall try to evaluate such commutators only 
at the limit $y \rightarrow x$. In analogy with the previous case,
we introduce a notation given by
$\rho_{i}(x',\lambda) 
\equiv \lim_{\epsilon\rightarrow 0+} \rho_i( x+\epsilon; \lambda )$. 
We are interested in calculating 
 commutators like 
 $\left[ \rho_{i}( x', \lambda ),
\psi( x ) \right] \equiv \lim_{\epsilon\rightarrow 0+} \left[
 \rho_{i}( x+\epsilon, \lambda ), \psi( x ) \right]$,
where $\epsilon\rightarrow 0+$ limit should be taken at the final stage
after evaluating all commutation relations. By using integral relations 
(3.6a,b) and canonical commutation relations (\ref {a4}) we obtain 
\bea
~~~~~~~~~~~&&\Big[ \rho_{1}( x', \lambda ), \psi( x ) \Big] \, = \, 
-i\hbar f \, \rho_{1}( x', \lambda ) \psi( x )
- i \hbar\xi\lambda \, \rho_{2}( x', \lambda ) \, , \nn ~~
 ~~~~~~~~~~~~~~~~~~(5.2a) \\
&&\left[ \rho_{1}( x', \lambda ), \psi^\dagger( x ) \right] \, = \,
i\hbar f\, \psi^\dagger( x )\rho_{1}( x', \lambda) \, ,
\nn \, ~~~~~~~~~~~~~~~~~~~~~~~~~~~~~~~~~~~~~~~~~(5.2b) \\
&&\Big[ \rho_{2}( x', \lambda ), \psi( x ) \Big] \, = \, i\hbar g
\, \rho_{2} ( x', \lambda )\psi( x ) \, ,
\nn ~~~~~~~~~~~~~~~~~~~~~~~~~~~~~~~~~~~~~~~~~~~~(5.2c) \\
&&\left[ \rho_{2}( x', \lambda ), \psi^\dagger( x ) \right] 
\, = \,  -i\hbar g \, \psi^\dagger(x)\rho_{2}( x', \lambda ) 
+ i\hbar \lambda\, \rho_{1}( x', \lambda ) \, . \nn
\, ~~~~~~~~~~~~~~~~~~~~(5.2d)  
\eea
\addtocounter{equation}{1}
The details of derivation for one of the above commutation
relations is given in Appendix A. 
It is clear from the relations (\ref {e1}) and (5.2) 
that the commutators 
 $\big[ \rho_i( y, \lambda ), \psi( x ) \big] $
and $\big[ \rho_i( y, \lambda ), \psi^\dagger(x) \big] $ are 
discontinuous at the point $y=x$. 
By repeatedly applying the commutation relations (5.2), we easily obtain 
\bea
~~~~~~~~~~&&\left[ \rho_{1}( x', \lambda),\psi^2(x) \right] \, = \, \hbar f
(\hbar f - 2i) \, \rho_1( x',\lambda )\psi^2(x)\nn \\ 
&&~~~~~~~~~~~~~~~~~~~~~~~~~~
-i\hbar\xi\lambda \, \Big \{ \, 2+i \hbar ( f-g ) \,  \Big\}
\, \rho_2( x', \lambda ) \psi( x ) \, ,\nn
 ~~~~~~~~~~~~~~~~(5.3a) \\
&&\left[ \rho_{1}( x', \lambda),{{\psi^\dagger}^2}(x) \right] \, = \,
 \hbar f (2i - \hbar f)
\, {{\psi^\dagger}^2}(x)\rho_1( x', \lambda ) \, ,\nn
\, ~~~~~~~~~~~~~~~~~~~~~~~~~~~~(5.3b)
\\
&&\left[ \rho_{2}( x', \lambda),\psi^2(x) \right]  \, = \, \hbar g
  ( 2i + \hbar g ) \, \rho_2( x',\lambda )\psi^2(x) \, ,
\nn~~~~~~~~~~~~~~~~~~~~~~~~~~~~~~~(5.3c)\\
&&\left[ \rho_{2}( x', \lambda),{{\psi^\dagger}^2}(x) \right] \, = \,
- \hbar g (2i + \hbar g) \,
{{\psi^\dagger}^2}(x)\rho_2( x', \lambda )\nn\\
&&~~~~~~~~~~~~~~~~~~~~~~~~~~~~+ i\hbar \lambda \, \Big\{ \, 2+i \hbar ( f-g )
 \,  \Big\} \,
\psi^\dagger( x )\rho_1( x', \lambda) \, .\nn  ~~~~~~~~~~~~~~~(5.3d) 
\eea
\addtocounter{equation}{1}

We would like to make a comment at this point. 
Since the integral relations 
of $\rho_{i}( x -\epsilon ,\lambda ) $
 and $\rho_{i}( x +\epsilon , \lambda ) $ 
coincide  with each other 
at the limit $\epsilon \rightarrow 0+$,
 one may say that the operators 
 $\rho_{i}( x', \lambda ) $ and $\rho_{i}( x'', \lambda ) $ are same 
in the `weak sense'.  However, we have already observed that the commutators 
 $\big[ \rho_i( y, \lambda ), \psi( x ) \big] $
and $\big[ \rho_i( y, \lambda ), \psi^\dagger(x) \big] $ are 
discontinuous at the point $y=x$. As a result,  operators of the form 
$\Delta_i(x,\lambda) \equiv \rho_i(x'',\lambda) - \rho_i(x',\lambda) $ 
yield nontrivial commutation relations with $\psi(x)$ and $\psi^\dagger(x)$. 
Thus, borrowing a terminology from the theory of constrained 
Hamiltonian systems [28], we may say that the operators 
 $\rho_{i}( x', \lambda ) $ and $\rho_{i}( x'', \lambda ) $ differ 
from each other in the `strong sense'. While deriving commutation
 relations like (5.2) in Appendix A, 
we have neglected some operators which become trivial in the weak sense 
at $\epsilon \rightarrow 0$ limit.  This procedure 
does not affect the validity of relations 
(5.2) in the weak sense. However, it is reasonable 
to ask whether the relations (5.2) are also valid in the strong sense. 
To investigate this point, one may try to evaluate 
commutators like 
$\left[ \rho_{i}( x', \lambda),\psi^2(x) \right] $ and 
$\left[ \rho_{i}( x', \lambda),{\psi^\dagger}^2(x) \right] $
 from the first principles.  This can be 
achieved with the help of integral relations (3.6a,b) and canonical 
 commutation relations (\ref {a4}), by evaluating at first the commutators 
$\left[ \rho_{i}( z, \lambda),\psi(y) \psi(x) \right] $ and 
$\left[ \rho_{i}( z, \lambda),\psi^\dagger(y) \psi^\dagger(x) \right] $ 
in the region 
$z>y>x$ and taking $y, \, z\rightarrow x$ limit at the 
final stage. One can verify that such a procedure will exactly reproduce 
the relations (5.3), which are obtained through repeated applications 
of the commutation relations (5.2). This fact suggests that the commutation
relations (5.2) are valid not only in the weak sense, but also in the
strong sense. 

Next, we try to evaluate commutation relations between basic field operators
and Jost solutions defined through boundary conditions at $x\rightarrow
+\infty$.  At first, we consider the commutators 
$\big[ \tau_i( y, \lambda ), \psi(x) \big] $ and 
$\big[ \tau_i( y, \lambda ),\psi^\dagger(x) \big] $
 in the region $y>x$.  For this case, all
   fields $\psi(z), \, \psi^\dagger(z)$ contained in 
the integral relations (3.13a,b)
 would commute with $\psi(x), \, \psi^\dagger(x)$. As a result, we get
 trivial relations like 
 $\Big[ \tau_i( y, \lambda ), \psi( x ) \Big] = \left[ 
\tau_i( y, \lambda ), \psi^\dagger(x) \right] = 0$ in the region
 $y > x $. The $y\rightarrow x$ limit of these commutation 
relations may be expressed in the form 
\bea
\Big[ \tau_{i}( x', \lambda ), \psi( x ) \Big] 
= \left[ \tau_{i}( x', \lambda ),
\psi^\dagger( x ) \right] = 0 \, ,
\label{e4}
\eea
where $\tau_{i}( x', \lambda ) \equiv \lim_{\epsilon\rightarrow 0+}
\tau_i(x+\epsilon ,\lambda)$.  Next, we consider the commutators 
 $\big[ \tau_i( y, \lambda ), \psi( x ) \big] $
and $\big[ \tau_i( y, \lambda ), \psi^\dagger(x) \big] $
 in the region $y<x$. However, it is difficult to find out
 these commutators in a closed form for arbitrary values of
 $x$ and $y$.  So, we shall evaluate such commutators only 
at the limit $y \rightarrow x$. By using integral relations 
(3.13a,b) and canonical commutation relations (\ref {a4}), we obtain 
\bea
~~~~~~~~~~~~~~~~~~&&\Big[ \tau_{1}( x'', \lambda ), 
\psi( x ) \Big] \, = \, i\hbar g\, \tau_{1}( x'', \lambda ) \psi( x )
+ i \hbar \xi\lambda \, \tau_{2}( x'', \lambda ) \, , \nn  ~~~~~~~~~~~~~~~
(5.5a) \\
&&\left[ \tau_{1}( x'', \lambda ), 
\psi^\dagger( x ) \right] \, = \,
 -i\hbar g\, \psi^\dagger(x)\tau_{1}( x'', \lambda ) \, , 
\nn \, ~~~~~~~~~~~~~~~~~~~~~~~~~~~~~~~(5.5 b) \\
&&\Big[ \tau_{2}( x'', \lambda ), 
\psi( x ) \Big] \, = \, -i \hbar f \, \tau_{2} (x'',\lambda)
\psi(x) \, , \nn \, ~~~~~~~~~~~~~~~~~~~~~~~~~~~~~~~~~(5.5c) \\
&&\left[ \tau_{2}( x'', \lambda ), 
\psi^\dagger( x ) \right] \, = \, 
i \hbar f \, \psi^\dagger(x)\tau_{2}(x'', \lambda ) 
- i\hbar\lambda \, \tau_{1}( x'', \lambda ) \, ,
\nn~~~~~~~~~~~~~~(5.5d)
\eea
\addtocounter{equation}{1}
where $\tau_{i}( x'', \lambda ) \equiv \lim_{\epsilon\rightarrow 0+}
\tau_i( x - \epsilon , \lambda )$. It is clear from the 
relations (\ref {e4}) and (5.5) that the commutators 
$\big[ \tau_{i}(y, \lambda ), \psi(x) \big] $ and 
$\big[ \tau_{i}(y, \lambda ), \psi^\dagger (x) \big] $
are discontinuous at the point $y=x$. 
By repeatedly applying the commutation relations (5.5), we also get
\bea
~~~~~~~~~~~&&\left[ \tau_1( x'', \lambda ), \psi^2(x) \right] \, = \,
\hbar g (2i + \hbar g) \, \tau_1( x'', \lambda ) \psi^2(x) \nn \\
&& \, ~~~~~~~~~~~~~~~~~~~~~~~~~~~+i\hbar \xi\lambda \,
\Big\{ \, 2+i\hbar ( f-g ) \, \Big \} \,
\tau_2( x'', \lambda )\psi( x ) \, ,\nn
 \, ~~~~~~~~~~~~~(5.6a) \\
&&\left[ \tau_1( x'', \lambda ), {\psi^\dagger}^2(x) \right] \, = \,
- \hbar g (2i+\hbar g){\psi^\dagger}^2(x) \, \tau_1( x'', \lambda ) \, ,\nn
 \, ~~~~~~~~~~~~~~~~~~~~~~~~~(5.6b)\\
&&\left[ \tau_2( x'', \lambda ), \psi^2(x) \right] \, = \, \hbar f
 (\hbar f - 2i) \, \tau_2(x'',\lambda )\psi^2(x) \, ,
\nn \, ~~~~~~~~~~~~~~~~~~~~~~~~~~~~~(5.6c) \\
&&\left[ \tau_2( x'', \lambda ), {\psi^\dagger}^2(x) \right] \, = \, 
 \hbar f (2i - \hbar f) \, 
{\psi^\dagger}^2(x)\tau_2( x'', \lambda )\nn \\
&&~~~~~~~~~~~~~~~~~~~~~~~~~~~~- i\hbar \lambda \, 
\Big \{ \, 2+i \hbar ( f-g ) \, \Big \} \, \psi^\dagger( x )
\tau_1( x'', \lambda ) \, .
~~~~~~~~~~~~~~(5.6d) \nn 
\eea
\addtocounter{equation}{1}

Till now we have derived all possible commutation relations between 
Jost solutions and basic field operators, which will be needed 
for our calculation of quantum conserved quantities.
Next, we consider commutation relations between two
 Jost solutions associated with different boundary conditions, i.e.
  commutators of the type 
$ \big [ \rho_i(y,\lambda) , \tau_j(x,\lambda) \big ] $ at the limit 
$y \rightarrow x$. By using the integral relations (3.6),
(3.13) and canonical commutation relations (\ref {a4}), 
it can be shown that 
$ \big [ \rho_i(x',\lambda) , \tau_j(x,\lambda) \big ] =
 \big [ \rho_i(x'',\lambda) , \tau_j(x,\lambda) \big ] =0$.
Thus, unlike the previous cases, the commutator 
$ \big [ \rho_i(y,\lambda) , \tau_j(x,\lambda) \big ] $
is continuous at the limit $y \rightarrow x$. Consequently, 
by following the method of extension [4], one may define the 
commutator $ \big [ \rho_i(x,\lambda) , \tau_j(x,\lambda) \big ] $
either as $ \big [ \rho_i(x',\lambda) , \tau_j(x,\lambda) \big ] $
or as $\big [ \rho_i(x'',\lambda) , \tau_j(x,\lambda) \big ]$. 
For both of these cases, one obtains the trivial result given by
\beq
 \Big [ \rho_i(x,\lambda) , \tau_j(x,\lambda) \Big ] = 0 \, .
\label {e7}
\eeq
Thus it is evident that, we can freely interchange 
the ordering of $\rho_i(x,\lambda)$ and $\tau_j(x,\lambda)$ 
in the expressions of quantum Wronskian (\ref {c17}) and 
  $ \Gamma_{\rho,\tau}(x, \lambda )$ operator (\ref {d1}). 

Next, we want to calculate the derivatives for 
bilinears of Jost solutions, i.e. quantities like
$\d_x  \, \big( \, \rho_i(x, \lambda)\tau_j(x, \lambda) \, \big) $. 
By using eqns.(\ref {c3}) and (\ref {c10}), it is easy to see that such a
derivative is given by the sum of few terms,
each of which is a product of Jost solutions and basic field
operators with arguments corresponding to exactly the same space point. 
 It is a standard practice [4,11] to express 
these terms in a form so that the operator 
$\psi^\dagger(x)$ ($\psi(x)$) is always placed at the extreme left 
(right), while the ordering of the remaining factors 
remains completely unchanged. 
For example, if the term $ \rho_i(x, \lambda) 
\psi(x) \psi^\dagger(x) \tau_j(x, \lambda)$ 
appears in a differential equation, it should be transformed to 
 $ \psi^\dagger(x) \rho_i(x, \lambda)
 \tau_j(x, \lambda) \psi(x) $.  For the purpose of
 expressing all terms 
in the above mentioned fashion, it is needed to use the 
 commutation relations between basic fields and Jost solutions 
associated with exactly same space point, i.e. commutators of the
form $[\rho_i(x,\lambda ),\psi(x)]$, $[\rho_i(x,\lambda ),\psi^\dagger (x)]$,
$[\tau_i(x,\lambda ),\psi(x)]$ and $[\tau_i(x,\lambda ),\psi^\dagger (x)]$.
We have commented earlier that,  evaluation of  
these commutators through integral relations (3.6) and (3.13) would  
lead to indeterminant integrals like 
$\int_{-\infty}^x \delta(x-z) F(z) dz $, where $F(z)$ is some function
of $z$. Similar to the case of NLS model [11], one may now try to fix 
these indeterminant integrals
 through a convention given by
 $\int_{-\infty}^x \delta(x-z) F(z) dz = {1\over 2} F(x)$. It can be 
easily checked that the above mentioned way of fixing 
indeterminant integrals and calculating $[\rho_i(x,\lambda ),\psi(x)]$, 
$[\rho_i(x,\lambda ),\psi^\dagger(x)]$ 
is essentially same as defining these commutators 
as 
\bea 
&&\Big [ \rho_i(x,\lambda ),\psi(x) \Big] ~\equiv~ {1\over 2} 
\Big [ \rho_i(x',\lambda ) + \rho_i(x'',\lambda ) ,\psi(x) \Big ] \, , \nn \\
&&\Big [ \rho_i(x,\lambda ),\psi^\dagger(x) \Big] ~\equiv~ {1\over 2} 
\Big [ \rho_i(x',\lambda ) + \rho_i(x'',\lambda ) ,\psi^\dagger(x) \Big] \, ,
\label {e8}
\eea
evaluating them through the relations
 (\ref {e1}) and (5.2) and 
 substituting the argument $x$ in place of $x'$ and $x''$ at the final stage.
  Similarly, one can calculate $[\tau_i(x,\lambda ),\psi(x)]$ and 
$[\tau_i(x,\lambda ),\psi^\dagger(x)]$,
by defining them exactly like 
(\ref {e8}) and using the relations (\ref {e4}) and (5.5).
 Explicit results for all of these commutation 
relations are given in Appendix B. 
However we find in Appendix B that, unlike the case of NLS model,  these 
commutation relations violate the Jacobi identity. 
Consequently, 
for the case of present DNLS model,
it is not meaningful 
to define commutation relations between Jost solutions and field operators 
with arguments at exactly same space point 
through the prescription (\ref {e8}). 

The above mentioned problem, which arises in the computation of 
$\d_x  \, \big( \, \rho_i(x, \lambda)\tau_j(x, \lambda) \, \big) $,
can be bypassed 
through the method of extension [4]. According to this method, the argument 
of one Jost solution is shifted by a small amount $\delta $ and 
$\delta \rightarrow 0$ limit is taken \emph {after} 
evaluating all relevant commutation relations.  The final result obtained 
in this way must be independent of the sign of $\delta $.  By applying this 
method of extension, and 
using differential equations (\ref{c3}),(\ref{c10}) as well as commutators 
(\ref{e1}), (5.2), (\ref{e4}), (5.5), we obtain 
\bea
&&\d_x
\Big ( \rho_1( x, \lambda )\tau_2( x, \lambda ) \Big )
 \, = \,  i\lambda\left\{ \, 
\xi \, \psi^\dagger( x )\rho_2( x, \lambda )\tau_2( x, \lambda ) 
+ \rho_1( x, \lambda )\tau_1( x, \lambda )\psi( x )  \, \right\}, \nn 
 \, ~~~~(5.9a) \\
%&&~~~~~~~~~~~~~~~~~~~~~~~~\nn \\
&&\d_x \Big(\rho_2( x, \lambda )\tau_1(x,\lambda)\Big)
\, = \, i\lambda\left\{ \,  \xi \, \psi^\dagger( x )\rho_2( x, \lambda )
\tau_2( x, \lambda ) + \rho_1( x, \lambda )
\tau_1( x, \lambda )\psi(x) \, \right\}, \nn
~~~~~(5.9b) \\
%&&~~~~~~~~~~~~~~~~~~~~~~~~~~~~~~~~~~~~\nn \\
&&\d_x \Big(\rho_1(x,\lambda)\tau_1(x,\lambda) \Big) 
\, = \,  -\frac{i\lambda^2}{2} \, \rho_1( x, \lambda )\tau_1( x, \lambda ) 
+ i (f+g) \,  \psi^\dagger(x)\rho_1( x, \lambda )\tau_1( x, \lambda )
\psi(x)\nn \\
&&~~~~~~~~~~~~~~~~~~~~~~~~~~~~~~~
+ i\xi\lambda \, \psi^\dagger(x)
\Gamma_{\rho,\tau}(x,\lambda) \, ,
\nn ~~~~~~~~~~~~~~~~~~~~~~~~~~~~~~~~~~~~~~~~~~(5.9c) \\
%&&~~~~~~~~~~~~~~~~~~~~~~~~~~~~~~~~~~\nn \\
&&\d_x \Big( \rho_2( x, \lambda )\tau_2( x, \lambda ) \Big)
\, = \,  \frac{i\lambda^2}{2} \, \rho_2( x, \lambda )\tau_2( x, \lambda ) -
i (f+g) \,  \psi^\dagger( x )\rho_2( x, \lambda )\tau_2( x, \lambda )
\psi(x)\nn \\
&&~~~~~~~~~~~~~~~~~~~~~~~~~~~~~~
+ i\lambda \, \Gamma_{\rho,\tau}(x,\lambda) \psi(x) \, . \nn
~~~~~~~~~~~~~~~~~~~~~~~~~~~~~~~~~~~~~~~~~~~~~~(5.9d)
\eea
\addtocounter{equation}{1}
Details of derivation for one of the above differential equations is given 
in Appendix C.  Using eqns.(\ref {c17}) and (5.9a,b),  we find that
\bea
\d_x \, \Lambda_{\rho,\tau}(x,\lambda) \,  = \,  0 \, .
\label{e10}
\eea
Thus we are able to explicitly show that, the quantum 
  Wronskian (\ref {c17}) remains independent of the variable $x$ 
even if the noncommutative nature of related 
operators are taken into account. 
Taking advantage of this fact,  we often 
use the notation $ \Lambda_{\rho,\tau}(\lambda) $, instead of 
$ \Lambda_{\rho,\tau}(x,\lambda) $, to denote the quantum Wronskian.
We are also interested in computing the derivative of 
$ \Gamma_{\rho,\tau}(x,\lambda) $ operator (\ref {d1}), since it appears 
in the r.h.s. of the fundamental relation (\ref {d2}).
With the help of eqns.(5.9a,b),  we easily obtain 
\bea
\d_x \Gamma_{\rho,\tau}(x,\lambda)
= 2i\lambda\left( \, \xi\psi^\dagger (x)
\rho_2( x, \lambda ) \tau_2( x, \lambda )
+ \rho_1( x, \lambda ) \tau_1( x, \lambda )
\psi(x) \, \right) \, .
\label{e11}
\eea
By using eqns.(\ref {e11}) and (5.9c,d), one can further show that 
\bea
\frac{\lambda}{4} \, \d_x \Gamma_{\rho,\tau}(x,\lambda)
-\frac{(f+g)}{2\lambda} \, \psi^\dagger(x)
\d_x \Gamma_{\rho,\tau}(x,\lambda) \, \psi( x )
\, = \, \Theta_{\rho,\tau}(x,\lambda) \, ,
\label{e12}
\eea
where
\bea
\Theta_{\rho,\tau}(x,\lambda) = \xi\psi^\dagger( x )
\d_x\Big( \,\rho_2( x, \lambda )
\tau_2( x, \lambda ) \,\Big) - \d_x \Big( \, \rho_1( x, \lambda )
\tau_1( x, \lambda ) \, \Big)\psi( x ).
\label{e13}
\eea

Finally, we try to find out 
commutation relations between the quantum Wronskian and 
basic field operators. 
 Since $\Lambda_{\rho, \tau}(y, \lambda )$ is shown to be independent of $y$, 
  commutators like $\Big[ \Lambda_{\rho, \tau} ( y, \lambda ), 
\psi( x ) \Big]$ and $\left[ \Lambda_{\rho, \tau}
(y, \lambda ), \psi^\dagger( x ) \right]$ 
should not depend on the choice of argument $y$. 
For the case of NLS model, 
such commutators are calculated for the choice $y=x$ [11].
However, we have already seen in Appendix B that this choice 
 leads to the violation of Jacobi identity for the case of DNLS model. 
So, instead of choosing $y=x$,
at present we shall calculate the commutators 
$\Big[ \Lambda_{\rho, \tau} ( y, \lambda ), 
\psi( x ) \Big]$ and $\left[ \Lambda_{\rho, \tau}
(y, \lambda ), \psi^\dagger( x ) \right]$ at the limit $y\rightarrow x $.
For this purpose, we introduce quantities like 
$\Lambda_{\rho,\tau}( x', \lambda ) \equiv \lim_{\epsilon\rightarrow 0+}
\Lambda_{\rho,\tau} ( x+\epsilon, \lambda )$ and 
$\Lambda_{\rho,\tau}( x'', \lambda ) \equiv 
\lim_{\epsilon\rightarrow 0+}\Lambda_{\rho,\tau}( x-\epsilon, \lambda )$.
Using eqns.(\ref {e1}), (5.2), (\ref {e4}) and (5.5), we find that the 
 commutators $\Big[ \Lambda_{\rho,\tau}( x', \lambda ), \psi( x ) \Big]$ 
 and $\Big[ \Lambda_{\rho,\tau}( x'', \lambda ), \psi( x ) \Big]$   
 yield the same result which may be expressed as
\bea
~~~~~~~~~~\Big[ \Lambda_{\rho,\tau}(\lambda ), \psi( x ) \Big] &=& -i\hbar f
\rho_1( x, \lambda )
\tau_2( x, \lambda )\psi( x ) - i\hbar g \, \rho_2( x, \lambda )\tau_1
( x, \lambda )\psi( x ) \nn \\ 
&-& i\hbar \xi\lambda \, \rho_2( x, \lambda )
\tau_2( x, \lambda ). \nn ~~~~~~~~~~~~~~~~~~~~~~~~~~~~~~
~~~~~~~~~~~~~~(5.14a)
\eea
In Appendix D we present the details for deriving the above relation.
Similarly, the commutators 
$\Big[ \Lambda_{\rho,\tau}(x',\lambda), \psi^\dagger (x) \Big]$ 
and $\Big[ \Lambda_{\rho,\tau}(x'',\lambda), \psi^\dagger (x) \Big]$ 
yield the same result given by
\bea
~~~~~~~~~~\left[ \Lambda_{\rho,\tau}(\lambda ), \psi^\dagger( x ) \right] 
&=& i\hbar f
\psi^\dagger( x )
\rho_1( x, \lambda )\tau_2( x, \lambda ) + i\hbar g\psi^\dagger( x )
\rho_2( x, \lambda )\tau_1( x, \lambda ) \nn \\ 
&-& i \hbar \lambda \rho_1( x, \lambda )
\tau_1( x, \lambda ). \nn ~~~~~~~~~~~~~~~~~~~~~~~~~~~~~~
~~~~~~~~~~~~~(5.14b)
\eea
\addtocounter{equation}{1}
Using eqns.(5.14a,b) and (5.9a-d), one can also find out the derivatives
of $\left[ \Lambda_{\rho,\tau}( \lambda ), \psi( x ) \right] $ and 
$\left[ \Lambda_{\rho,\tau}(\lambda ), \psi^\dagger(x)\right]$ as
\bea
~~~~~\partial_x\Big[ \Lambda_{\rho,\tau}( \lambda ), 
\psi( x ) \Big] &=& \hbar\lambda (f+g) \, 
\rho_1(x,\lambda) \tau_1(x,\lambda) \psi^2(x)
\nn \\
&+& \frac{\hbar \xi\lambda^3}{2}\rho_2( x, \lambda )\tau_2( x, \lambda ) 
+ \hbar \xi\lambda^2  \, \Gamma_{\rho,\tau}(x,\lambda) \psi( x )~~~\nn \\ 
&-&i\hbar \, \Big( f \, \rho_1( x, \lambda ) \tau_2( x, \lambda ) 
+g \, \rho_2( x, \lambda )\tau_1 ( x, \lambda )\Big ) \, \d_x\psi(x) \, ,  
~~~~~~~~(5.15a)\nn
\eea
and
\bea
~~~~
\d_x\left[ \Lambda_{\rho,\tau}( \lambda ), \psi^\dagger( x ) \right]
&=& -\hbar\lambda\xi (f+g) \, {\psi^\dagger}^2(x)
\rho_2( x, \lambda ) \tau_2( x, \lambda )
\nn \\
&-& \frac{\hbar \lambda^3}{2}\rho_1(x, \lambda) \tau_1( x, \lambda ) 
+ \hbar \xi\lambda^2 \, \psi^\dagger(x) \Gamma_{\rho,\tau}(x,\lambda) 
\nn \\ 
&+& i\hbar \,
\partial_x\psi^\dagger(x)\, \Big( f\rho_1( x, \lambda )\tau_2( x, \lambda )
+ g\rho_2( x, \lambda )
\tau_1( x, \lambda )\Big) \, .
\, ~~~~~~~(5.15b)\nn
\eea
\addtocounter{equation}{1}
We are further interested in evaluating commutation relations between 
$\Lambda_{\rho,\tau} (y,\lambda)$ and the square of basic field operators.
Proceeding as before, it is shown in Appendix D that the commutators 
$\Big[ \Lambda_{\rho,\tau}( x', \lambda ),
\psi^2(x) \Big]$ and $\Big[ \Lambda_{\rho,\tau}( x'', \lambda ), 
\psi^2(x) \Big]$ yield the same result given by
\bea
~~~~~~~~\Big[ \Lambda_{\rho,\tau}(\lambda ), \psi^2(x) \Big]&=& 
\hbar f\left(\hbar f -2i\right)\rho_1( x, \lambda )\tau_2( x, \lambda )
\psi^2(x) \nn \\ 
&~&-\hbar g\left(2i +\hbar g \right) \,
\rho_2( x, \lambda )\tau_1( x, \lambda )\psi^2(x) \nn\\
&~& - i\hbar\xi\lambda
\Big\{ \, 2 + i \hbar ( f-g ) \, \Big\} \, \rho_2( x, \lambda )
\tau_2( x, \lambda )\psi(x) \, . \nn~~~~~~~~~~~~~~~
(5.16a) \nn 
\eea
Similarly, the commutators 
$\left[ \Lambda_{\rho,\tau}(x',\lambda ), {\psi^\dagger}^2(x) \right]$
and $\left[ \Lambda_{\rho,\tau}(x'',\lambda ), {\psi^\dagger}^2(x) \right]$
yield
\bea
~~~~~~~\left[ \Lambda_{\rho,\tau}(\lambda ), {\psi^\dagger}^2(x) \right]&=&
- \hbar f \left( \hbar f -2i \right) \, {\psi^\dagger}^2(x)\rho_1(x,\lambda)
\tau_2( x, \lambda ) \nn \\ 
&~& + \hbar g\left( 2i +\hbar g \right) \, {\psi^\dagger}^2(x)\rho_2(x,\lambda)
\tau_1(x,\lambda)\nn\\
&~& - i\hbar \lambda \, \Big\{ 2+i\hbar ( f-g ) \Big\}\, \psi^\dagger( x )
\rho_1( x, \lambda )\tau_1
( x, \lambda ) \, .\nn ~~~~~~~~~~~~~~~(5.16b)\nn
\eea
\addtocounter{equation}{1}
All of these relations will be extensively used in our calculation 
of quantum conserved quantities for the DNLS model.
 
\vspace{1cm}

\noindent \section 
{Explicit Construction of the Quantum Hamiltonian and its
spectrum}
\renewcommand{\theequation}{6.{\arabic{equation}}}
\setcounter{equation}{0}

\medskip
Here we try to find out the explicit form of the first few
 quantum conserved quantities of DNLS model,  which would satisfy the 
fundamental relation (\ref {d2}).
Analogous to the classical case (1.3a), we take the first quantum 
conserved quantity to be
\bea
I_0 = \int_{-\infty}^{+\infty}\psi^\dagger( x )\psi( x ) dx \, .
\label{f1}
\eea
Using (5.14a,b), we find that 
\bea
&&\Big[ \Lambda_{\rho,\tau}( \lambda ), I_0 \Big] = \int_{-\infty}^{+\infty}
\left\{ \,
\left[ \Lambda_{\rho,\tau}( \lambda ), \psi^\dagger( x ) \right]\psi( x ) +
\psi^\dagger( x )\Big[ \Lambda_{\rho,\tau}( \lambda ), \psi( x ) 
\Big] \, \right\} dx \nn \\
&&\, ~~~~~~~~~~~~~~~= -i\hbar \lambda   \int_{-\infty}^{+\infty}
\left\{ \rho_1 \tau_1\psi( x ) + \xi \psi^\dagger( x )
\rho_2\tau_2 \right\} dx \, .
\label {f2}
\eea
Note that, in the above relation and in the rest of this section, we omit
the arguments of Jost solutions $\rho_i(x, \lambda)$ and
 $\tau_i(x, \lambda)$ for the sake of convenience.
With the help of (\ref {e11}), equation (\ref {f2}) can be simplified as 
\bea
\Big[ \Lambda_{\rho,\tau}( \lambda ), I_0 \Big] 
= -\frac{\hbar}{2}\int_{-\infty}^{+\infty}
\d_x \Gamma_{\rho,\tau}(x,\lambda) \, dx =
-\frac{\hbar}{2} \left[ \, \Gamma_{\rho,\tau}( +\infty, \lambda ) 
- \Gamma_{\rho,\tau} (-\infty, \lambda) \, \right].
\label{f3}
\eea
So one concludes that for $n=0$, the fundamental relation
(\ref{d2}) is satisfied by $I_0$.

By imitating its classical counterpart (1.3b), 
the second quantum conserved quantity may be taken as
\bea
I_1 = - i\int_{-\infty}^{+\infty}\psi^\dagger( x )
\partial_x\psi(x) \, dx \, .
\label{f4}
\eea
Neglecting some integrals of total derivatives which lead to vanishing 
surface terms, one can write the commutation relation between 
$\Lambda_{\rho,\tau}(\lambda)$ and $I_1$ (\ref {f4}) as
\[
\Big[\Lambda_{\rho,\tau}(\lambda), I_1 \Big] = i\int_{-\infty}^{+\infty}
\left\{ \, \d_x \left[ \Lambda_{\rho,\tau}( \lambda ), 
\psi^\dagger(x) \right] \, \psi(x) 
- \psi^\dagger(x) \, \d_x\Big[ \Lambda_{\rho,\tau}
( \lambda ), \psi( x ) \Big] \, \right\} dx \, . 
\]
Applying further (5.15a,b) and neglecting 
 some integrals of total derivatives, we find that 
\bea
\Big[\Lambda_{\rho ,\tau}(\lambda), I_1\Big] 
&=& \hbar \int_{-\infty}^{+\infty}
 \Big \{ - \frac{i \lambda^3}{2} 
\left( \xi \psi^\dagger(x) \rho_2 \tau_2 +
\rho_1 \tau_1 \psi(x) \right) \nn \\
&~&~~~~~~~~~~~ + \psi^\dagger(x) \Big( f \d_x \left( \rho_1\tau_2 \right) 
+ g \d_x \left( \rho_2\tau_1 \right) 
\Big) \psi(x) \nn \\
&~&~~~~~~~~~~~ -i \lambda (f+g) \psi^\dagger(x) 
\left( \xi\psi^\dagger(x)\rho_2\tau_2
+ \rho_1 \tau_1 \psi(x) \right) \psi(x) \Big \} \, . \nn
\eea
Using (5.9a,b) and (\ref {e11}) to simplify the r.h.s.
of above relation, we readily obtain 
\bea
\Big[ \Lambda_{\rho,\tau}( \lambda ), I_1 \Big] = -\frac{\hbar\lambda^2}{4}
\int_{-\infty}
^{+\infty}\frac{\partial \Gamma_{\rho,\tau}(x,\lambda)}{\partial x} dx
= -\frac{\hbar\lambda^2}{4}\left[ \, \Gamma_{\rho,\tau}( +\infty, \lambda ) - 
\Gamma_{\rho,\tau}( -\infty, \lambda ) \, \right].
\label{f5}
\eea
Thus $I_1$ satisfies the fundamental relation (\ref{d2}) for $n=1$.

Finally, we try to calculate the quantum Hamiltonian of DNLS model.
In analogy with its classical counterpart (1.3c), we propose that 
this quantum Hamiltonian can be written in the form 
\bea
I_2 = I_2^{( 1 )} +  i \xi_q  I_2^{( 2 )} \, ,
\label{f6}
\eea
where
\bea
~~~~~~~~~~I_2^{( 1 )} = - 
\int_{-\infty}^{+\infty} \psi^\dagger( x ) \partial_{xx} \psi(x) \, dx \, ,
~~~ I_2^{( 2 )} = \int_{-\infty}^{+\infty} {\psi^\dagger}
^2( x )\partial_x \psi^2( x ) \,  dx \, , \nn ~~~~~~~~~~~(6.7a,b )
\eea
\addtocounter{equation}{1}
and $\xi_q$ is some yet undetermined coupling constant. 
 Neglecting some integrals of total derivatives
   which lead to vanishing surface terms,
one can write the commutation relation between 
$\Lambda_{\rho,\tau}(\lambda)$ and $I_2^{(1)}$ (6.7a) as 
\[
\left[ \Lambda_{\rho,\tau}( \lambda ), I_2^{( 1 )} \right] = 
\int_{-\infty}^{+\infty}\left
\{ \, \partial_x
\left[ \Lambda_{\rho,\tau}( \lambda ), \psi^\dagger( x )  \right] 
\partial_x\psi( x ) + 
\partial_x\psi^\dagger( x )\partial_x\Big[  \Lambda_{\rho,\tau}(\lambda ), 
\psi( x )  \Big]
\, \right\} dx \, .
\]
Using (5.15a,b) to evaluate the commutators appearing
in the r.h.s of above relation and neglecting again
  integrals of some total derivatives, we obtain
\bea
\left[ \Lambda_{\rho,\tau}( \lambda ), I_2^{( 1 )} \right]&=&
\hbar\int_{-\infty}^{+\infty} \Big[ \,
-\xi\lambda^2\psi^\dagger(x)
\partial_x \Gamma_{\rho,\tau}(x,\lambda)\psi(x)
+ \lambda (f+g)
\psi^\dagger(x)\Theta_{\rho,\tau}(x,\lambda)\psi(x) \nn \\
&-&\frac{\lambda^3}{2}\Theta_{\rho,\tau}(x,\lambda) 
+ 2\lambda (f+g) \psi^\dagger(x) \Big(\,\xi
\partial_x\psi^\dagger(x)\rho_2 \tau_2 - \rho_1 \tau_1
\partial_x\psi( x ) \,\Big)\psi(x)\,\Big] dx \, , \cr 
&~&
\label {f8} 
\eea 
where $\Theta_{\rho,\tau}(x,\lambda) $ is given by (\ref {e13}).
Using the identity (\ref {e12}) and substituting explicit values 
of $f$ and $g$ (i.e., $f = \xi
e^{-i\alpha/2} / (\cos \alpha / 2)$ , $g = \xi
e^{i\alpha/2} /(\cos \alpha / 2) \, $),
 equation (\ref {f8}) can be written in the form 
\bea
\left[ \Lambda_{\rho,\tau}( \lambda ), I_2^{( 1 )} \right] &=&
\hbar\int_{-\infty}^{+\infty}\Big[ \, 
- \frac{\lambda^4}{8} \d_x \Gamma_{\rho,\tau}(x,\lambda)
- 2{\xi}^2{\psi^\dagger}^2(x) \d_x \Gamma_{\rho,\tau}(x,\lambda)
 \psi^2(x) \nn \\ 
&~&~~~~~~~~+ 4\lambda \xi \, \psi^\dagger(x) \Big(\,\xi
\partial_x\psi^\dagger(x)\rho_2 \tau_2 - \rho_1 \tau_1
\partial_x\psi( x ) \,\Big)\psi(x)
 \, \Big]  dx \, . 
\label{f9} 
\eea

Next, we consider the commutation relation between 
$\Lambda_{\rho,\tau}(\lambda)$ and $I_2^{(2)}$ (6.7b). 
 Neglecting the integral of a total derivative, we can write this 
commutator as
\bea
\left[ \Lambda_{\rho,\tau}( \lambda ), I_2^{( 2 )} \right] 
= \int_{-\infty}^{+\infty}
\left\{ \, \left[ \Lambda_{\rho,\tau}( \lambda ), 
{\psi^\dagger}^2( x ) \right] \partial_x\psi^2( x )
- \partial_x{\psi^\dagger}^2(x)
\left[ \Lambda_{\rho,\tau}( \lambda ), \psi^2( x ) \right]
\, \right\} dx. \nn 
\eea
Applying (5.16a,b), 
 neglecting again integrals of some total derivatives,  and also
using relations like $\d_x(\rho_1 \tau_2) = 
\d_x(\rho_2 \tau_1) = \frac{1}{2} \d_x\Gamma_{\rho,\tau}(x,\lambda)$,
the above equation can be brought in the form 
\bea
&&\left[ \Lambda_{\rho,\tau}( \lambda ), I_2^{( 2 )} \right]=
\frac{\hbar}{2}
\int_{-\infty}^{+\infty}\Big[ \, ( \, \hbar f^2-\hbar g^2-2if-2ig \, )
{\psi^\dagger}^2( x )
\partial_x \Gamma_{\rho,\tau}(x,\lambda)\psi^2( x )~~~~~~~~~~~~~~ \nn \\
&&~~~~~~~~~~~~~~~+ 4i\lambda \big( 2+i\hbar ( f-g ) \big)
\psi^\dagger(x) \Big(\,\xi
\partial_x\psi^\dagger(x)\rho_2 \tau_2 - \rho_1 \tau_1
\partial_x\psi( x ) \,\Big)\psi(x) \, \Big] dx \, .
\label{f10} 
\eea
Using eqns.(\ref {f9}), (\ref {f10}) (with explicit values of
$f,~g$)  and (\ref {a6}), we find that 
the quantum Hamiltonian (\ref {f6}) would satisfy the fundamental
relation given by
\bea
&&\Big[ \Lambda_{\rho,\tau}( \lambda ), I_2 \Big] 
= -\frac{\hbar\lambda^4}{8}
\int_{-\infty}^{+\infty}\frac{\partial \Gamma_{\rho,\tau}(x,\lambda)}
{\partial x } dx
= -\frac{\hbar\lambda^4}{8} \left[ \, \Gamma_{\rho,\tau}( +\infty, \lambda ) 
- \Gamma_{\rho,\tau}( -\infty, \lambda ) \, \right] ,\nn\\
&~&~~~~~~~~~~~~~~~~~~~~~~~~~~~
\label {f11}
\eea
provided the parameter $\xi_q$ is chosen as
\beq
\xi_q = \frac {\xi}{\sqrt {1- \hbar^2\xi^2}} \, .
\label {f12}
\eeq
By substituting (\ref{f12}) in (\ref{f6}), we get 
an explicit expression for the quantum Hamiltonian of DNLS model as 
\bea
I_2 = \int_{-\infty}^{+\infty}
\left\{ - \psi^\dagger(x) \d_{xx} \psi(x) 
+  \frac {i\xi}{\sqrt {1- \hbar^2\xi^2}} \, 
{\psi^\dagger}^2(x) \partial_x \psi^2(x) \right\}
dx \, .
\label{f13}
\eea
Thus, it is established that the above quantum Hamiltonian satisfies 
the fundamental relation (\ref {d2}) for $n=2$. 
Even though this quantum Hamiltonian (\ref {f13}) is not manifestly Hermitian, 
we can easily make it Hermitian by adding some
  integrals of total derivatives which lead to vanishing surface terms.
Comparing (\ref {f13}) with (1.3c) we surprisingly find that,
due to quantum effect, the coupling constant of the system is modified. 
Consequently, unlike most other integrable systems,  
the quantum Hamiltonian of DNLS model can not be obtained from its
classical counterpart by simply applying the normal ordering prescription. 
It is interesting to note that, eqn.(\ref {f12}) is somewhat 
similar to the relation between rest mass and dynamical mass
of a relativistic particle given by: 
$m = \frac {m_o}{\sqrt {1- v^2/c^2}} $, where $m_0$, $m$ and $v/c$
play the role of $\xi$, $\xi_q$ and $\hbar \xi$ respectively. 
The $v/c \rightarrow 0$ limit is like $\hbar \rightarrow 0$
limit (for a fixed $\xi$) in our case. Just as the dynamical mass
of a relativistic particle coincide with its rest mass in 
 the nonrelativistic limit, the quantum 
coupling constant $\xi_q$ (\ref {f12}) coincides with the bare coupling
constant $\xi$ at $\hbar \rightarrow 0$ limit. On the other hand, 
the $v/c \rightarrow 1$ limit is analogous to
$ \v \xi \v \rightarrow {1\over \hbar}$ limit in our case. 
 Just as the dynamical mass of a particle goes to infinity at 
ultrarelativistic limit, $\xi_q$ (\ref {f12}) goes to infinity 
at $ \v \xi \v \rightarrow {1\over \hbar}$ limit. Consequently, 
even though QYBE restricts the value of $\xi$ as 
 $ \v \xi \v  \leq \frac{1}{\hbar}$, there exists no such restriction 
on the value of corresponding quantum coupling constant $\xi_q$ (\ref {f12}). 
Thus the apparent limitation about the applicability of 
QISM in solving quantum DNLS Hamiltonian for the full range of its 
coupling constant is resolved in a very nice way.

It is evident that $I_0$ (\ref {f1}) and $I_1$ (\ref {f4}) represent 
 the number operator and momentum operator  respectively for the 
quantum DNLS system. Substituting $n=0,$ $1$ and $2$ in 
equation (\ref {d7}), one can explicitly write down 
 the eigenvalue relations for $I_0$, $I_1$ and $I_2$ as
\bea
&&~~~~~~~~~~~~~~~I_0 \, \v \mu_1,\mu_2,\cdots ,\mu_N \r \, = \, \hbar N \,
\v \,  \mu_1,\mu_2,\cdots ,\mu_N \r \, , \nn  
\, ~~~~~~~~~~~~~~~~~~~~~~~~~~~~~~~(6.14a) \\
&&~~~~~~~~~~~~~~~I_1 \, \v \mu_1,\mu_2,\cdots ,\mu_N \r \, = \,
 \Big( \, \frac{\hbar}{2}\sum_{j=1}^N \mu_j^{2} \, \Big) \, \v
 \,  \mu_1,\mu_2,\cdots ,\mu_N \r \, , \nn 
~~~~~~~~~~~~~~~~~~~~~~(6.14b) \\
&&~~~~~~~~~~~~~~~I_2 \, \v \mu_1,\mu_2,\cdots ,\mu_N \r \, = \,
 \Big( \, \frac{\hbar}{4}\sum_{j=1}^N \mu_j^{4} \, \Big) \v \,
 \,  \mu_1,\mu_2,\cdots ,\mu_N \r \, . \nn 
 \, ~~~~~~~~~~~~~~~~~~~~~~(6.14c)
\eea
\addtocounter {equation}{1}
Let us now compare these 
  eigenvalue relations with those obtained 
through the technique of coordinate Bethe ansatz. Projecting 
the bosonic Hamiltonian (\ref {f13}) on an $N$-particle Hilbert space [22],
we get
\beq
{\cal H}_N = - \hbar \sum_{j=1}^N \frac{\d^2}{\d x_j^2} + 
2i \hbar^2 \xi_q \sum_{l<m} \delta(x_l -x_m) 
\left( \frac{\d}{\d x_l} + \frac{\d}{\d x_m}  \right) \, .
\label {f15}
\eeq
The eigenvalues for this Hamiltonian with derivative $\delta$-function 
interaction and corresponding 
momentum operator can be derived through the method 
of coordinate Bethe ansatz [22,23]. It is easy to check that such 
eigenvalues completely match with our result in eqns.(6.14b,c) 
when we identify the momentum parameters ($k_j$) of 
coordinate Bethe ansatz with the spectral parameters ($\mu_j$) 
of present approach through the relation: 
$k_j \equiv \frac {\mu_j^2}{2}$. The eigenfunctions of 
the Hamiltonian (\ref {f15}) can also be constructed through 
coordinate Bethe ansatz. If, for the simplest $N=2$ case,  such 
eigenfunction is chosen in $x_1<x_2$ region as 
$f(x_1, x_2) = e^{i(k_1 x_1 + k_2 x_2)} $, then its form in 
 $x_1>x_2$ region would be given by [22,24]
\[
f(x_1, x_2) = A(k_1,k_2) e^{i(k_1 x_1 + k_2 x_2)} 
+ B(k_1, k_2) e^{i(k_2 x_1 + k_1 x_2)} \, 
\]
where 
$A(k_1, k_2) =  \frac {k_1 -k_2 + i\hbar \xi_q (k_1 + k_2 )}{k_1-k_2}$ and 
$B(k_1,k_2) =  1- A(k_1,k_2)$ are the so called `matching coefficients'. 
With the help of these matching coefficients, one can easily find out the 
$S$-matrix for two-body scattering as [24]
\beq
S(k_1,k_2) = A(k_1,k_2) A(k_2,k_1)^{-1} = \frac {k_1-k_2 
+ i\hbar \xi_q (k_1+k_2)}
{k_1-k_2 - i\hbar \xi_q (k_1+k_2)} \, .
\label {f16}
\eeq
Using eqns.(\ref {f12}) and (\ref {a6}), we can express $\xi_q$ as: 
 $\xi_q = -\frac{1}{\hbar}\tan \alpha$.
Putting this form of $\xi_q$ in eqn.(\ref {f16}), and identifying 
momentum parameters with spectral parameters through relations like
$k_1\equiv \frac {\lambda^2}{2}, ~k_2\equiv \frac {\mu^2}{2}$, we find 
that this $S$-matrix (\ref {f16})
 exactly matches with our earlier result (\ref {b13})
which is derived in the framework of QISM. The fact that
the `renormalized' coupling constant $\xi_q$  appears in the 
projected DNLS Hamiltonian (\ref {f15}), instead of its classical
counterpart $\xi$, plays a crucial role in this
 comparison between the results of coordinate
and algebraic Bethe ansatz. 

It is also interesting to compare the results of coordinate
and algebraic Bethe ansatz 
for the soliton sector of quantum DNLS model. 
By applying QISM it is found that, 
 the distribution of complex spectral 
parameters for such  quantum $N$-soliton state is given by 
the relation (\ref {b19}). Taking the square of both sides of this 
relation and substituting 
$k_j $ in place of $ \mu_j^2/2$, we obtain 
\beq
  k_j ~=~ \frac{\mu^2}{2}
 \, \exp \left[  i \alpha \left( N+1 -2j \right)
\right] \, ,
\label {f17}
\eeq
where $j\in [1,2, \cdots N]$. This equation coincides with the 
 momentum distribution in coordinate Bethe ansatz corresponding to
  the quantum $N$-soliton states of DNLS Hamiltonian (\ref {f13}) [22,23].  
Again, the fact that the modified coupling constant appears in the 
  Hamiltonian (\ref {f13}) allows us to exactly
match the results of coordinate and algebraic Bethe ansatz.

By using the eigenvalue relations (6.14b,c), 
we can also calculate the binding energy for the above mentioned 
quantum $N$-soliton states. Substituting the values of 
complex $\mu_j$ (\ref {b19}) to (6.14b), we obtain the momentum 
eigenvalue corresponding to these $N$-soliton states as
\beq
P= \frac{\hbar\mu^2}{2}\sum_{j=1}^N e^{i\alpha(N+1-2j)} 
=\frac{\hbar\mu^2\sin{\alpha N}}{2\sin \alpha} \, .
\label{f18}
\eeq
Similarly, by substituting 
 $\mu_j$ (\ref {b19}) to (6.14c), we obtain the energy
eigenvalue corresponding to these states as
\bea
E = \frac{\hbar\mu^4}{4}\sum_{j=1}^N e^{2i\alpha(N+1-2j)} 
=\frac{\hbar\mu^4\sin(2\alpha N)}{4\sin(2\alpha)} \, .
\label{f19}
\eea
To calculate binding energy, we assume that the momentum $P$ 
(\ref {f18}) of the
$N$-soliton state is equally distributed among $N$ number of 
single-particle scattering states. The real (pure imaginary)
 spectral parameter associated with each of these 
single-particle states is denoted by $\mu_0$. 
With the help of eqns.(6.14b) and (\ref{f18}), we obtain 
\beq
\mu_0^2=\frac{\mu^2\sin(\alpha N)}{N\sin \alpha}. 
\label{f20}
\eeq
Using eqn.(6.14c), one can easily calculate 
the total energy for 
 $N$ number of such single-particle scattering states as
\bea
E'= \frac{\hbar N}{4}\mu_0^4 
= \frac{\hbar\mu^4\sin^2\alpha N}{4N\sin^2 \alpha} \, .
\label{f21}
\eea
Subtracting  $E$ (\ref {f19}) from $E'$ (\ref {f21}), we obtain the 
binding energy of quantum $N$-soliton state as 
\beq
E_B = E'- E 
= \frac{\hbar\mu^4\sin \alpha N}{4\sin \alpha}\Big\{\frac{
\sin \alpha N}{N\sin \alpha} - \frac{\cos \alpha N}{\cos \alpha} \Big\} .
\label{f22}
\eeq
Substituting $N=2$ to the above relation,
we obtain $E_B =\frac{\hbar\mu^4}{2}\sin^2 \alpha$. Thus we get
$E_B >0$ for any nonzero value of $\alpha$. 
For N=3, (\ref{f22}) takes the form
$E_B = \frac{2\hbar\mu^4}{3}\sin^2\alpha(3-4\sin^2\alpha)$.
Here we get $E_B>0$ only if $\v \alpha \v <\frac{\pi}{3}.$
Applying the method of induction, we find that the condition 
 $E_B>0$ is in fact valid within the range $\v \alpha \v <\frac{\pi}{N}$
for all values of $N$ [29]. Thus to obtain quantum $N$-soliton states
with positive binding energy, the coupling constant of DNLS model 
should be restricted within the region $\v \xi_q \v < {1\over \hbar}
\tan \left( \frac{\pi}{N} \right)$. 

\medskip

\noindent \section {Concluding Remarks}

In analogy with the `fundamental relation' of NLS model [11], in this article 
we  propose the fundamental relation (\ref {d2}) for DNLS model. This 
fundamental relation plays a key role in our 
construction of quantum conserved quantities of DNLS model and their 
spectra.  However, from the technical point of view,
our construction of quantum conserved quantities
is much more complicated than the case of NLS model due to the following 
reasons.  Quantum Jost solutions and their commutation relations 
with basic field operators are extensively used 
to obtain the quantum conserved quantities of DNLS model.
It turns out that, in contrast to the case of NLS model,
 differential equations satisfied by these Jost solutions 
corresponding to boundary conditions at $x \rightarrow \infty$ and 
$x \rightarrow - \infty$ do not coincide with each other. This 
salient feature of DNLS model is connected with the fact that
its quantum Lax operator (\ref {a5}) has a nonvanishing trace.
We also find that, unlike the case of NLS model,
 the commutation relation between Jost solutions of DNLS model 
and basic field operators with arguments at exactly the same space 
point lead to the violation of Jacobi identity.
So we are compelled to use commutation relations between Jost solutions 
and basic field operators with slightly shifted arguments in our calculation 
of quantum conserved quantities.

Proceeding in the above mentioned way, we are able to explicitly
construct the quantum Hamiltonian and few other 
conserved quantities of DNLS model through basic field 
operators of this system. Surprisingly we find that, 
unlike the cases of most other integrable systems, this quantum 
Hamiltonian (\ref {f13})  can not be obtained as normal ordered version of the 
corresponding classical Hamiltonian (1.3c). This is due to the fact that 
 a new kind of coupling constant ($\xi_q$),  quite 
different from the classical one ($\xi$), appears in the quantum
Hamiltonian of the DNLS model. 
Thus we obtain the explicit form of the quantum Hamiltonian of 
DNLS model, which has been defined earlier
in the framework of QISM in a formal way.
Interestingly, the relation (\ref {f12})
between $\xi$ and $\xi_q$ is rather similar to the relation 
between rest mass and dynamical mass of a relativistic particle. 
 Just as the dynamical mass
of a relativistic particle coincides with its rest mass in 
 the nonrelativistic limit, 
 $\xi_q$  coincides with $\xi$ at $\hbar \rightarrow 0$ limit. 
In the ultrarelativistic limit, 
 the dynamical mass of a particle tends towards infinity. 
In a similar way, $\xi_q$ 
can take arbitrary large value at 
$ \v \xi \v  \rightarrow \frac{1}{\hbar}$ limit.
Consequently, we can apply
QISM to the quantum DNLS model for the full range of its coupling
constant,  even though QYBE restricts the value of $\xi$ as 
 $ \v \xi \v  \leq \frac{1}{\hbar}$.
Due to the presence of modified coupling constant
in the quantum Hamiltonian (\ref {f13}),
we are also able to consistently match various results of algebraic and 
coordinate Bethe ansatz in the case of DNLS model. The $S$-matrix 
for two particle scattering and the distribution of 
single-particle momentum for 
quantum $N$-soliton states are two such examples where the results 
of algebraic and coordinate Bethe ansatz match with each other.
We also calculate the binding energy for the quantum $N$-soliton 
state of DNLS model and find out the range of coupling constant 
for which this binding energy has a positive value. 

As a future study, it might be interesting to find out the higher quantum 
conserved quantities of DNLS model by using its fundamental relation and 
investigate whether the coupling constants appearing in such higher 
conserved quantities also differ from their classical counterparts. 
It is well known that,   higher quantum 
conserved quantities of NLS model can not be expressed in normal ordered 
form as the integral of a one-dimensional density [12,13]. 
A similar situation may 
also arise for the case of higher quantum conserved quantities of the 
DNLS model.  It should be noted that, the present approach of using 
 fundamental relation for the construction of quantum conserved quantities 
 and their spectra might be applicable to many other integrable systems. 
In this article, we have already discussed the possibility of 
such construction for a class of quantum integrable field models 
associated with $2\times 2$ Lax equations (\ref {c19}). 
It should also be interesting to
study fundamental relations for the case of discrete quantum integrable
 models like  Heisenberg spin-$\frac{1}{2}$ chain, supersymmetric t-J 
model, Hubbard model etc. and explore how these
 fundamental relations lead to the construction of corresponding 
 conserved quantities along with their spectra.

\medskip

%\noindent {\bf Acknowledgments }

\newpage
\leftline {\large \bf Appendix A}

\medskip
Here we give a detailed derivation of the commutation relation (5.2a). 
At first, we shall evaluate the commutator 
$\Big[ \rho_1( x+ \epsilon , \lambda ), \psi( x ) \Big]$ for the case
$\epsilon >0$ and  take $\epsilon \rightarrow 0$ limit at the final
stage.  Using the integral relation of $\rho_1( x, \lambda )$ (3.6a) and
canonical commutation relations (\ref {a4}), we find that
\bea
\Big[ \rho_1( x+\epsilon, \lambda ), \psi( x ) \Big] &=&
if \int_{-\infty}^{x+\epsilon}dz \, e^{\frac{i\lambda^2}{4}( z-x-\epsilon )}
\Big[ \psi^\dagger( z ), \psi( x ) \Big]
\rho_1( z, \lambda )\psi( z ) \nn \\
&+&if\int_{-\infty}^{x+\epsilon}dz \, e^{\frac{i\lambda^2}{4}( z-x-\epsilon )}
\psi^\dagger( z )\Big[ \rho_1( z, \lambda ), \psi( x ) \Big]
\psi( z ) \nn \\
&+& i\xi\lambda\int_{-\infty}^{x+\epsilon}dz \, e^{\frac{i\lambda^2}{4}
( z-x-\epsilon )}\left[ \psi^\dagger( z ), \psi( x ) \right]
\rho_2( z, \lambda )\nn
\\
&+&i\xi\lambda\int_{-\infty}^{x+\epsilon}dz \, e^{\frac{i\lambda^2}{4}
( z-x-\epsilon )}\psi^\dagger( z )\Big[ \rho_2( z, \lambda ), 
\psi( x ) \Big]\nn
\\
&=& -i\hbar f e^{\frac{i\lambda^2}{4}
( -\epsilon )}\rho_1( x, \lambda )\psi( x )
- i\hbar\xi\lambda e^{\frac{i\lambda^2}{4}( -\epsilon )}\rho_2( x, \lambda )
+ \Omega. \nn ~~~~~~~~~~(A1)
\eea
where
\bea
~~~~~~~~~~~~~
\Omega &=& i f \int_x^{x+\epsilon}dz \, 
e^{\frac{i\lambda^2}{4}( z-x-\epsilon )}
\psi^\dagger( z )\Big[ \rho_1( z, \lambda ), \psi( x ) \Big]
\psi( z ) \nn \\
&+& i\xi\lambda\int_x^{x+\epsilon}dz \,
 e^{\frac{i\lambda^2}{4}( z-x-\epsilon )}
\psi^\dagger( z )\Big[ \rho_2( z, \lambda ), \psi( x ) \Big]. \nn 
~~~~~~~~~~~~~~~~~~~~~~~~~~~~~~ (A2)
\eea
The lower limits of integrals appearing in the r.h.s. of eqn.(A2)
are fixed by using the fact that the commutator 
$\big[ \rho_i(z,\lambda), \psi(x) \big]$ becomes trivial for the case
$z <x$. Next, we rewrite eqn.(A1) as
\bea
&& \Big[ \rho_1( x+\epsilon, \lambda ), \psi( x ) \Big] 
= -i\hbar f  e^{\frac{i\lambda^2}{4}
(-\epsilon)}\rho_1( x+\epsilon, \lambda )\psi( x ) - i\hbar\xi\lambda 
e^{\frac{i\lambda^2}{4}( -\epsilon )}\rho_2( x+\epsilon, \lambda ) 
+ \Omega + \Omega' , ~~\nn \\
&&~~~~~~~~~~~~~~~~~~~~~~~~~~~~~~~~~~~~~~~~~~~~~~~~~~~~~~~~~~~~~~~
~~~~~~~~~~~~~~~~~~~~~~~~~~~~~~~~~~~~~~~~~~~(A3) \nn
\eea
where
\bea
\Omega' = -i\hbar f e^{\frac{i\lambda^2}{4}( -\epsilon )}
\Big[ \rho_1( x, \lambda )-
\rho_1( x+\epsilon, \lambda )\Big]\psi( x ) 
- i\hbar\xi\lambda  \, 
e^{\frac{i\lambda^2}{4}( -\epsilon )}\Big[ \rho_2( x, \lambda ) - \rho_2( 
x+\epsilon, \lambda)\Big]. \nn
\eea
It is clear that the above expression of 
$\Omega'$ vanishes at $\epsilon\rightarrow 0$ limit. 
Let us now assume that commutators like  
$\Big[ \rho_i(z , \lambda ), \psi( x ) \Big]$  do not produce any 
singular term at the limit $z\rightarrow x$. Due to this assumption,
 the operator $\Omega $ (A2) 
 would also vanish at $\epsilon\rightarrow 0$ limit. 
Consequently, by taking $\epsilon \rightarrow 0$ limit of (A3),
we obtain the commutation relation (5.2a). Other commutation 
relations appearing in (5.2) can also be derived in a similar fashion. 
It should be noted that,
the forms of finally derived equations (5.2) justify
in a self-consistent way our assumption
about the absence of singular terms in commutators like 
$\Big[ \rho_i(z , \lambda ), \psi( x ) \Big]$ at 
 $z\rightarrow x$ limit.

\newpage

\leftline {\large \bf Appendix B}
\medskip

Here we derive the
commutation relations between Jost solutions and field operators 
associated with the same space point through the
prescription (\ref {e8}) and show that these commutation relations 
violate the Jacobi identity.
Inserting the commutators 
(\ref {e1}) and (5.2) to the expression (\ref {e8}), 
and substituting the arguments $x'$ and $x''$ by $x$ at the final stage, 
we find that
\bea
~~~~~~~~~~~&&\Big[ \rho_{1}( x, \lambda ), \psi( x ) \Big] \, = \, 
-\frac{i\hbar f}{2} \, \rho_{1}( x, \lambda ) \psi( x )
- \frac{i\hbar\xi\lambda}{2} \, \rho_{2}( x, \lambda ) \, , \nn ~~
\, ~~~~~~~~~~~~~~~~~~(B1.1) \\
&&\left[ \rho_{1}( x, \lambda ), \psi^\dagger( x ) \right] \, = \,
\frac{i\hbar f}{2} \, \psi^\dagger( x )\rho_{1}( x, \lambda) \, ,
\nn \, ~~~~~~~~~~~~~~~~~~~~~~~~~~~~~~~~~~~~~~~~~(B1.2) \\
&&\Big[ \rho_{2}( x, \lambda ), \psi( x ) \Big] \, = \, \frac{i\hbar g}{2}
\, \rho_{2} ( x, \lambda )\psi( x ) \, ,
\nn ~~~~~~~~~~~~~~~~~~~~~~~~~~~~~~~~~~~~~~~~~~~~(B1.3) \\
&&\left[ \rho_{2}( x, \lambda ), \psi^\dagger( x ) \right] 
\, = \, - \frac{i\hbar g}{2} \, \psi^\dagger(x)\rho_{2}( x, \lambda ) 
+ \frac{i\hbar\lambda}{2} \, \rho_{1}( x, \lambda ) \, . \nn
\, ~~~~~~~~~~~~~~~~~~~~(B1.4)  
\eea
Similarly, one can calculate the commutators 
$[\tau_i(x,\lambda ),\psi(x)]$ and 
$[\tau_i(x,\lambda ),\psi^\dagger(x)]$,  by defining them 
exactly like (\ref {e8}) and using the relations (\ref {e4}) 
as well as (5.5). In this way, we obtain
\bea
~~~~~~~~~~~&&\Big[ \tau_{1}( x, \lambda ), 
\psi( x ) \Big] \, = \, \frac{i\hbar g}{2} \, \tau_{1}( x, \lambda ) \psi( x )
+ \frac{i\hbar\xi\lambda}{2}
 \, \tau_{2}( x, \lambda ) \, , \nn \, ~~~~~~~~~~~~~~~~~~~~~~~ (B2.1) \\
&&\left[ \tau_{1}( x, \lambda ), 
\psi^\dagger( x ) \right] \, = \,
 -\frac{i\hbar g}{2} \, \psi^\dagger(x)\tau_{1}( x, \lambda ) \, , 
\nn ~~~~~~~~~~~~~~~~~~~~~~~~~~~~~~~~~~~~~~~(B2.2) \\
&&\Big[ \tau_{2}( x, \lambda ), 
\psi( x ) \Big] \, = \, - \frac{i \hbar f}{2} \, \tau_{2} (x,\lambda)
\psi(x) \, , \nn ~~~~~~~~~~~~~~~~~~~~~~~~~~~~~~~~~~~~~~~~~(B2.3) \\
&&\left[ \tau_{2}( x, \lambda ), 
\psi^\dagger( x ) \right] \, = \, 
\frac{i \hbar f}{2} \, \psi^\dagger(x)\tau_{2}(x, \lambda ) 
- \frac{i\hbar \lambda}{2} \, \tau_{1}( x, \lambda ) \, .
\nn~~~~~~~~~~~~~~~~~~~~~~~(B2.4)
\eea
Due to eqn.(\ref {e7}),  Jost solutions $\rho_i(x,\lambda )$
and $\tau_j(x,\lambda )$ commute with each other. 

By successively using the
 commutators (B1.1) and (B2.3), we find that 
\bea
~~~~~~~~~~~~~~~
\Big [\tau_2(x, \lambda),\big[ \rho_{1}( x, \lambda ), \psi( x ) \big] \Big]
 \, = \,  -\frac{\hbar^2 f^2}{4}
 \rho_{1}(x,\lambda) \tau_{2}(x,\lambda) \psi(x) \, . \nn 
~~~~~~~~~~~~~~~~~~~~~(B3)
\eea
Next, by applying the commutators 
(B2.3), (B1.1) and (\ref {e7}), we obtain
\bea
~~~\Big [\rho_1(x, \lambda),\big[ \psi(x), \tau_2( x, \lambda ) \big] \Big] 
 \, = \, \frac{\hbar^2 f^2}{4}
 \rho_{1}(x,\lambda) \tau_{2}(x,\lambda) \psi(x) 
+ \frac{\hbar^2 f \xi \lambda}{4} 
  \rho_{2}(x,\lambda) \tau_{2}(x,\lambda) \, . \nn  \, ~~~(B4)
\eea
Finally, by using  eqns.(B3), (B4) and (\ref {e7}), it is easy to check 
that
\bea
&&\Big [\tau_2(x, \lambda),\big[ \rho_{1}( x, \lambda ), \psi( x ) \big] \Big]
+ \Big [\rho_1(x, \lambda),\big[ \psi(x), \tau_2( x, \lambda ) \big] \Big] 
+ \Big [ \psi(x) ,\big[\tau_2( x, \lambda ), \rho_1(x, \lambda) \big] \Big] 
\nn \\
&&~~~~~~~~~~~~~~~~~~~~~~~~~~~~~~~~~~~~~~~~~~~~~~~~~~~~~~~~~~=~ 
 \frac{\hbar^2 f \xi \lambda}{4} 
  \rho_{2}(x,\lambda) \tau_{2}(x,\lambda) \, . \nn ~~~~~~~~~~~~(B5)
\eea
Thus it is evident that the set of commutation relations (B1), (B2) and 
(\ref {e7}) violate the Jacobi identity. 

\medskip
\newpage

\leftline {\large \bf Appendix C}

\medskip
For deriving the relation (5.9a) through the method of 
extension, we shift the argument of $\rho_1(x, \lambda)$
by a very small amount $ \delta $ and find out
 $\d_x\Big( \rho_1( x +\delta , \lambda ) 
\tau_2( x, \lambda ) \Big)$ for both positive
and negative $\delta$. For both cases, $\delta \rightarrow 0$ limit 
 will be taken at the final stage.  It will be shown
that the final result is independent of the sign of $\delta $.

Let us first take a positive $\delta $. 
Using eqns.(\ref{c3}) and 
(\ref {c10}) we get
\bea
\d_x\Big(\,\rho_1( {\tilde x}, \lambda )
\tau_2( x, \lambda ) \,\Big)
&=&  \, \d_x\rho_1( {\tilde x} , \lambda ) \,  \tau_2(x, \lambda ) 
+ \rho_1( {\tilde x}, \lambda )\, \d_x \tau_2(x,\lambda) \,  \nn \\
&=&\Big\{  if \, \psi^\dagger({\tilde x}) 
\rho_1({\tilde x}, \lambda )\psi({\tilde x}) 
+ i\xi\lambda \, \psi^\dagger({\tilde x})
\rho_2({\tilde x},\lambda) \Big \} \, \tau_2( x, \lambda )  \nn \\
&+& \rho_1({\tilde x}, \lambda ) \, 
\Big \{ -if \, \psi^\dagger(x)\tau_2(x, \lambda )\psi( x ) + i\lambda \,
\tau_1( x, \lambda )\psi( x ) \Big \} \,  , ~~~~(C1) \nn
\eea
where $ {\tilde x}\equiv x+\delta $. The 
 r.h.s. of (C1) should be written in a way such that an
 operator like $\psi^\dagger(x) $ ($ \, \psi(x) \, $) 
is always placed at the extreme left (right) 
of each term.  The $\delta 
\rightarrow 0$ limit should be taken after rewriting
the r.h.s of (C1) in the above mentioned way with the help of 
commutators $\left [ \psi({\tilde x}) , \tau_2(x,\lambda) \right ]$ 
and  $\left [ \rho_1( {\tilde x}, \lambda ) , \psi^\dagger(x) \right ]$.
Thus we have to ultimately use
 the $\delta \rightarrow 0$ limit of 
commutators $\left [ \psi({\tilde x}) , \tau_2(x,\lambda) \right ]$ 
and  $\left [ \rho_1({\tilde x}, \lambda ) , \psi^\dagger(x) \right ]$,
which are given by eqns.(5.5c) and (5.2b) respectively. 
By using these equations and dropping 
terms which vanish at $ \delta \rightarrow 0$ limit, we obtain
\bea
&& \d_x \Big( \,\rho_1( {\tilde x }, \lambda )\tau_2( x, \lambda
) \,\Big) \nn \\
&& ~~~~=if( \, 1+i\hbar f \, )\Big[ \, \psi^\dagger({\tilde x} )
\rho_1( {\tilde x}, \lambda )\tau_2( x, \lambda )\psi({\tilde x}) 
- \psi^\dagger( x )\rho_1({\tilde x}, \lambda )\tau_2( x,\lambda)
\psi(x) \, \Big] \nn \\
&&~~~~~~ + i\xi\lambda \, \psi^\dagger({\tilde x})
\rho_2({\tilde x}, \lambda )\tau_2( x,
\lambda ) + i\lambda \, \rho_1({\tilde x}, \lambda )\tau_1( x, \lambda )
\psi( x ) \nn \\
&&~~~~=  i\xi\lambda \, \psi^\dagger(x)
\rho_2(x,\lambda)\tau_2(x,\lambda) 
+ i\lambda \, \rho_1(x,\lambda)\tau_1(x,\lambda) \psi(x) \,  . \nn 
~~~~~~~~~~~~~~~~~~~~~~~~~~~~(C2)
\eea

Next, we consider the case $\delta < 0 $. In this case also 
  we obtain a relation of the form (C1), where $ \tilde x\equiv x+\delta $.
Again we want to rewrite the r.h.s. of (C1) in a way such that an
 operator like $\psi^\dagger(x) $ ($ \, \psi(x) \, $) 
is always placed at the extreme left (right) 
of each term.  However it is already found in Sec.5 that,
  $\left [ \psi({\tilde x}), \tau_2(x,\lambda) \right ]$ =
  $\left [ \rho_1({\tilde x},\lambda) , \psi^\dagger(x) \right ]=0$
for any negative value of $\delta$.  By using these commutation relations 
and dropping terms which vanish at $ \delta \rightarrow 0$ limit, 
we obtain
\bea
 \d_x \Big( \,\rho_1( {\tilde x }, \lambda )\tau_2( x, \lambda
) \,\Big) &=& if\Big[ \, \psi^\dagger({\tilde x} )
\rho_1( {\tilde x}, \lambda )\tau_2( x, \lambda )\psi({\tilde x}) 
- \psi^\dagger( x )\rho_1({\tilde x}, \lambda )\tau_2( x,\lambda)
\psi(x) \, \Big] \nn \\
&+& i\xi\lambda \, \psi^\dagger({\tilde x})
\rho_2({\tilde x}, \lambda )\tau_2( x,
\lambda ) + i\lambda \, \rho_1({\tilde x}, \lambda )\tau_1( x, \lambda )
\psi( x ) \nn \\
&=&  i\xi\lambda \, \psi^\dagger(x)
\rho_2(x,\lambda)\tau_2(x,\lambda) 
+ i\lambda \, \rho_1(x,\lambda)\tau_1(x,\lambda) \psi(x) \,  . \nn 
~~~~~~~(C3)
\eea

Comparing the r.h.s. of (C2) and (C3),
we find that $\d_x \Big( \rho_1
(x,\lambda)\tau_2( x, \lambda )\Big)$ is given by
eqn.(5.9a) in a regularisation independent way.

\medskip
\newpage 

\leftline {\large \bf Appendix D}

\medskip
For the purpose of deriving the relation (5.14a), we write down 
$\Lambda(x',\lambda)$ and $\Lambda(x'',\lambda)$ explicitly as 
\bea
~~~~~~~~~~&&\Lambda( x', \lambda ) = \rho_1( x', \lambda )\tau_2
( x', \lambda ) - \rho_2( x', \lambda )\tau_1( x', \lambda ) \, ,
\nn~~~~~~~~~~~~~~~~~~~~~~~~~~~~~~~~(D1) \\
~~~~~~~~~~&&\Lambda( x'', \lambda ) = \rho_1( x'', \lambda )\tau_2
( x'', \lambda ) - \rho_2( x'', \lambda )\tau_1( x'', \lambda ) \, .
\nn~~~~~~~~~~~~~~~~~~~~~~~~~~~~~ \, (D2)
\eea

Using (D1), (5.2a), (5.2c) and (\ref{e4}), we find that 
\bea
\Big[ \Lambda(x',\lambda), \psi(x) \Big] &=& \Big[ \rho_1
( x', \lambda ), \psi(x) \Big]\tau_2( x', \lambda ) - 
\Big[ \rho_2 (x',\lambda), \psi(x) \Big]\tau_1( x', \lambda ) \nn \\
&=& - i\hbar f\rho_1( x', \lambda )\tau_2( x', \lambda )\psi( x ) -
i\hbar g \rho_2( x', \lambda )\tau_1( x', \lambda )\psi( x )\nn \\
&~& - i\hbar \xi\lambda\rho_2( x', \lambda )\tau_2( x', \lambda ) \, .\nn
~~~~~~~~~~~~~~~~~~~~~~~~~~~~~~~~~~~~~~~~~~~~~~~~~~(D3)
\eea
Similarly, using (D2), (\ref{e1}), (5.5a) and (5.5c), we get 
\bea
\Big[ \Lambda( x'', \lambda ) , \psi(x) \Big] &=&
\rho_1( x'', \lambda )\Big[\tau_2( x'', \lambda ) , \psi(x) \Big] -
\rho_2( x'', \lambda )\Big[ \tau_1( x'', \lambda ), \psi(x) \Big] \nn \\
&=& - i\hbar f\rho_1( x'', \lambda )\tau_2( x'', \lambda )\psi( x ) -
i\hbar g\rho_2( x'', \lambda )\tau_1( x'', \lambda )\psi( x ) \nn \\
&~& - i\hbar \xi\lambda\rho_2( x'', \lambda )\tau_2( x'', \lambda ) \, . \nn
~~~~~~~~~~~~~~~~~~~~~~~~~~~~~~~~~~~~~~~~~~~~~~~~~(D4)
\eea
Comparing (D3) and (D4), we find that 
$\big[ \Lambda(x',\lambda), \psi(x) \big] $ and 
$\big[ \Lambda(x'',\lambda), \psi(x) \big] $ 
lead to the same result (in the weak sense)
given by eqn.(5.14a). 

Next, we want to derive the relation (5.16a). Using eqns.(D1), (5.3a), (5.3c)
and (\ref {e4}), we obtain
\bea                                      
\Big[ \Lambda( x', \lambda ) , \psi^2(x) \Big] &=& 
\Big[ \rho_1( x', \lambda ) , \psi^2(x) \Big]\tau_2( x', \lambda ) -
\Big[ \rho_2( x', \lambda ) , \psi^2(x) \Big]\tau_1( x', \lambda ) \nn \\
&=& \hbar f ( \hbar f -2i) 
\rho_1( x', \lambda )\tau_2( x', \lambda )\psi^2( x )
 \nn \\
&~&- \hbar g ( 2i + \hbar g)
\rho_2( x', \lambda )\tau_1( x', \lambda )\psi^2( x ) \nn \\
&~&- i \hbar \xi\lambda\Big( \, 2+i \hbar ( f-g )\,\Big)\rho_2( x', \lambda )
\tau_2( x', \lambda ) \psi(x) \, .\nn
~~~~~~~~~~~~~~~~~~~~~~(D5) 
\eea
Similarly, using (D2), (\ref{e1}), (5.6a) and (5.6c), we get 
\bea
\Big[\Lambda(x'',\lambda) , \psi^2(x)\Big] &=&
\rho_1(x'',\lambda)\Big[ \tau_2(x'',\lambda), \psi^2(x) \Big]
-\rho_2( x'', \lambda)\Big[ \tau_1(x'', \lambda) , \psi^2(x) \Big] \nn \\
&=&  \hbar f( \hbar f -2i) 
\rho_1( x'', \lambda )\tau_2( x'', \lambda )\psi^2( x ) \nn \\
&~&- \hbar g ( 2i + \hbar g)
\rho_2( x'', \lambda )\tau_1( x'', \lambda )\psi^2( x ) \nn \\
&~&- i \hbar\xi\lambda\Big( \, 2+i \hbar( f-g )\, \Big)\rho_2( x'', \lambda )
\tau_2( x'', \lambda ) \psi(x) \, .\nn
~~~~~~~~~~~~~~~~~~~~~(D6) 
\eea
Comparing (D5) and (D6), again we find that 
$\big[ \Lambda(x',\lambda), \psi^2(x) \big] $ and 
$\big[ \Lambda(x'',\lambda), \psi^2(x) \big] $ lead to the same result
given by eqn.(5.16a). 

\newpage

\leftline {\large \bf References}
\medskip
\begin{enumerate}
\item P. Fendley, A.W.W. Ludwig and H. Saleur, Phys. Rev. Lett. 74 (1995)
     3005; P. Fendley, H. Saleur and N.P. Warner, 
     Nucl. Phys. B 430 (1994) 577; P. Fendley, A.W.W. Ludwig and H. Saleur, 
     Phys. Rev. B 52 (1995) 8934. 

\item G. Montambaux, D. Poliblanc, J. Bellisard and C. Sire, 
Phys. Rev. Lett. 70 (1993) 497; 
 D. Poliblanc, T. Ziman,  J. Bellisard, F. Mila and 
 G. Montambaux, Europhys. Lett. 22 (1993) 537. 

\item L.D. Faddeev, Sov. Sci. Rev. C1 (1980) 107; in {\it Recent Advances in
Field Theory and Statistical Mechanics, } ed. J.B.Zuber and R.Stora ,
 (North-Holland, Amsterdam, 1984 ) p.561. 

\item  E.K. Skylanin, in  
Yang-Baxter Equation in Integrable systems, Advanced series in Math. Phys. 
Vol. 10, edited by M. Jimbo ( World Scientific, Singapore, 1990) p.121.

\item E.K. Skylanin, L.A. Takhtajan and L.D. Faddeev, Theor. Math. Phys.
 40 (1980) 688.

\item V. E. Korepin, N. M. Bogoliubov, and A. G. Izergin, 
{\it Quantum Inverse 
Scattering Method and Correlation Functions} (Cambridge Univ. Press, 
Cambridge, 1993) and references therein. 

\item H.B. Thacker, Physica D 18 (1986) 348.

\item K. Sogo and M. Wadati, Prog. Theor. Phys. 69 (1983) 431.

\item F.H.L. Essler and V.E. Korepin, Phys. Rev. B 46 (1992) 9147.

\item J. Links, H.Q. Zhou, R.H. McKenzie and M.D. Gould, 
      Phys. Rev. Lett. 86 (2001) 5096. 

\item K. M. Case, J. Math. Phys. 25 (1984) 2306.

\item M. Omote, M. Sakagami, R. Sasaki and I. Yamanaka, Phys. Rev. D 35 
      (1987) 2423.

\item  B. Davies, Physica A 167 (1990) 433.

\item M. Wadati and A. Kuniba, J. Phys. Soc. Jpn. 55 (1986) 76.

\item D.J. Kaup and A.C. Newell, J. Math. Phys. 19 (1978) 798.

\item H.H. Chen, Y.C. Lee and C.S. Liu, Phys. Scr. 20 (1979) 490.

\item M. Wadati, H. Sanuki, K. Konno and Y.H. Ichikawa, 
      Rocky Mountain J. Math. 8 (1978) 323; Y.H. Ichikawa and 
      S. Watanabe, J. de Physique 38 (1977) C6-15; Y.H. Ichikawa, 
      K. Konno, M. Wadati and H. Sanuki, J. Phys. Soc. Jpn. 48 (1980) 279. 

\item P. A. Clarkson, Nonlinearity 5 (1992) 453.

\item A. Kundu and B. Basu-Mallick, J. Math. Phys. 34 (1993) 1052.

\item B. Basu-Mallick and A. Kundu, Phys. Lett. B 287 (1992) 149.

\item B. Basu-Mallick and T. Bhattacharyya, Nucl. Phys. B 634 [FS] (2002) 611.

\item A.G. Shnirman, B.A. Malomed and E.B. Jacob, Phys. Rev. A 50 (1994)
      3453.

\item D. Sen, {\it Quantization of the derivative nonlinear 
      Schr${\rm {\ddot o}}$dinger equation}, \hfil \break
cond-mat/9612077.

\item E. Gutkin, Ann. Phys. 176 (1987) 22.

\item E. Gutkin, Phys. Rep. 167 (1988) 1.

\item H.B. Thacker and D. Wilkinson, Phys. Rev. D 19 (1979) 3660; 
      D.B. Creamer, H.B. Thacker and
      D. Wilkinson, Phys. Rev. D 21 (1980) 1523. 

\item A.B. Zamolodchikov and A.B. Zamolodchikov, Ann. Phys. 120 (1979) 253. 

\item P.A.M. Dirac, {\it Lectures on quantum mechanics}, Belfer Graduate 
      School Monograph Series No. 2, Yeshiva University (1964).

\item B. Basu-Mallick , T. Bhattacharyya and D. Sen, {\it Bound and anti-bound
  soliton states for a quantum integrable derivative nonlinear 
  Schr${\ddot {\rm o}}$dinger model}, hep-th/0305252.

\end{enumerate}

\end{document}